\documentclass[usenatbib,onecolumn]{mn2e}
\usepackage{txfonts}
\usepackage{epsfig}
\usepackage{lscape}
\usepackage{graphicx}

\def\newline{\hfil\break}

\def\apj{{ApJ}}                 % Astrophysical Journal
\def\apjl{{ApJ}}                % Astrophysical Journal, Letters
\def\apjs{{ApJS}}               % Astrophysical Journal, Supplement
\def\mnras{{MNRAS}}             % Monthly Notices of the RAS
\def\nat{{Nature}}              % Nature
\title[Constraining Dynamical Dark Energy Models from BH Abundance]{Constraining Dynamical Dark Energy Models through 
the Abundance of High-Redshift Supermassive Black Holes}
\author[Lamastra et al.]{A. Lamastra$^1$\thanks{E-mail: alessandra.lamastra@oa-roma.inaf.it}, N. Menci$^1$, F. Fiore$^1$, C. Di Porto$^{2,3,4}$, L. Amendola$^3$\\
$^1$ INAF - Osservatorio Astronomico di Roma, via di Frascati
33, I-00040 Monteporzio, Italy\\
$^2$ Dipartimento di Fisica “E. Amaldi”, Universit\`a degli Studi “Roma Tre”, 
via della Vasca Navale 84, 00146, Roma, Italy\\
$^3$Institut f\"ur Theoretische Physik, Universit\"at Heidelber, Philosophenweg 16, D-69120 Heidelberg, Germany\\
$^4$ INAF-Osservatorio Astronomico di Bologna, Via Ranzani 1, Bologna 40127, Italy
}
\date{Accepted 2011 November 15. Received 2011 November 07; in original form 2011 February 14}
\pagerange{\pageref{firstpage}--\pageref{lastpage}} \pubyear{}

\begin{document}

\maketitle
\begin{abstract}
We compute the number density of massive Black Holes (BHs) at the centre of galaxies at $z=6$ in different Dynamical Dark Energy (DDE) cosmologies, and compare it with existing observational lower limits, to derive constraints on the evolution of the Dark Energy equation of state parameter $w$. Our approach only assumes the canonical scenario for structure formation from the collapse of overdense regions of the Dark Matter dominated primordial density field on progressively larger scales; the Black Hole accretion and merging rate have been maximized in the computation so as to obtain robust constraints on $w$ and on its look-back time  derivative $w_a$.
Our results provide independent constraints complementary to those obtained by combining Supernovae, Cosmic Microwave Background  and Baryonic Acoustic Oscillations; while the latter concern combinations of $w_0$ and $w_a$ leaving the time evolution of the state parameter $w_a$ highly unconstrained, the BH abundance mainly provide upper limits on $w_a$, only weakly depending on $w_0$. Combined with the existing constraints, our results significantly restrict the allowed region in DDE parameter space, ruling out DDE models not providing cosmic time and fast growth factor large enough to allow for the building up of the observed abundance of BHs; in particular, models with $-1.2\leq w_0\leq -1$ and positive redshift evolution $w_a\gtrsim 0.8$ - completely consistent with previous constraints -  are strongly disfavoured  by our independent constraints from BH abundance. Such range of parameters corresponds to "Quintom" DDE models, with $w$ crossing $-1$ starting from larger values. 
%We show how determining the duty cycle for Black Hole accretion could greatly tighten the constraints on the classes of DDE models with equation of state parameter increasing with redshift $w_a>0$; while hints in  this direction can be provided by detailed models of galaxy formation including a detailed description of Black Hole accretion, we show how such a goal can be achieved  by future observations of the shape of the high-redshift quasar luminosity function. 
\end{abstract}


\section{Introduction}

The measurements of distances of  Type Ia SNae on cosmological scales (up to $z\approx 1$; Riess et al. 1998; Perlmutter et al. 1999), when interpreted in the standard Friedmann-Lemaitre cosmological models, imply that the Universe 
expansion is accelerating. When such measurements are compared with those concerning the power spectrum of the Cosmic Microwave Background (CMB; see Dunkley et al. 2009), the correlation length of large scale-structures (Eisenstein et al. 2005), or the abundance of galaxy clusters (Bahcall et al. 2003), they indicate that 70 \% 
of the energy density in the Universe is constituted by some sort of negative pressure field (named Dark Energy DE). 
Its contribution $\Omega_{DE}\approx 0.7$ (in units of the critical density) results to be 
larger (but of the same order at present time) than the energy density $\Omega_M \approx 0.27$ contributed by Dark Matter (DM), and much larger than that 
from ordinary baryonic matter $\Omega_b\approx 0.03$. Although such results can be accommodated 
in a "Concordance Cosmology" (Spergel et al. 2007) consistent with a wide set of observations, they pose an enormous challenge for fundamental physics, when interpreted in terms of vacuum energy density (for a critical discussion see Bianchi, Rovelli, Kolb 2010 and references therein). 

In fact, while the proposed extensions of the particle physics standard model provide natural candidates for the DM, 
the nature of DE constitutes a real conundrum. The  simplest candidate is the Einstein cosmological constant
(see Carrol, Press, Turner 1992; Weinberg 1989; Sahni, Starobinsky 2000), characterized by constant energy density $\rho$ and pressure $p$, yielding an equation of state $w\equiv p/\rho=-1$. However, when interpreted in terms of vacuum energy density, 
the value required to induce the observed acceleration would be unnaturally small $\rho\approx (10^{-3}\,eV)^4$  compared to the natural energy scales in particle physics; besides, the coincidence problem of why this value is at present close to the DM energy density has no natural explanation (Zlatev, Wang, Steinhardt 1999). 
The problems related to the above fine tuning have led to consider the general 
case where the DE equation-of-state ratio $w(a)$ varies with time (here expressed in terms of the Universe scale factor $a$). Such a {\it dynamical} Dark Energy (DDE) is indeed expected if 
the DE results from the pressure $p_{\phi}=\dot \phi-V(\phi)$ and the energy density $\rho_{\phi}=\dot \phi+V(\phi)$ of a scalar field $\phi$ (named Quintessence, see Peebles, Ratra 1988; Caldwell, Dave, Steinhardt 1998; Sahni, Wang, 2000; Copeland, Sahni,  Tsujikawa,  2006; Frieman, Turner, Huterer 2008 ) evolving in a potential $V(\phi)$.  A time-varying $w(a)$ also provides a phenomenological rendition of modifications in the Friedmann equations for the rate of expansion (see Linder, Jenkins 2003) due to alternative theories of gravitation.

Characterizing the DE equation of state would constitute a fundamental step toward understanding the nature of DE, with a great impact on fundamental physics, involving quantum theory, high-energy physics, gravitational physics and astrophysics. However, at present  the time evolution of the equation of state is not very well constrained. Recent constraints come from: i) Type Ia Supernovae. These are based on measurements of SNae luminosity-distances up to $z\approx 1$ (see, e.g., Davis et al. 2007); ii)  CMB power spectrum, based on the measurements of the angular-diameter distance corresponding to the baryonic acoustic scale at the time of recombination ($z=1089$); iii) the baryonic acoustic oscillations observed in the correlation function of galaxies at $z\approx 0.35$, providing a ruler to measure the ratio of the distances to $z=0.35$ and $z=1089$ (Eisenstein et al. 2005). 
Further constraints on DDE could be provided by the weak lensing tomography, i.e., the measurement of the distortion in the images of distant source galaxies due to the inhomogeneous mass distribution  of the Universe, over a range of cosmic times (Massey et al. 2007). The size of these distortions depends upon both the distances travelled, and 
upon the growth function $D(a)$ which determines the amplitude of the deflector mass concentrations (Amara \& Refregier 2007; see Refregier 2003 for a review).

Combining the above measurements still leaves allowed a relatively large volume of the parameter space for the 
local value and the time variation of the equation-of-state parameter (see, e.g, Linder \& Jenkins 2003; Komatsu et al. 2008; Li, Xia, Fan, Zhangm 2008; Kowalski et al. 2008); parametrizing its evolution as  $w(a)=w_0+w_a(1-a)$ (an expression shown by Linder 2003 to provide an excellent approximation to a wide variety of scalar fields and other DE models), present observations yield 
$w_0\approx -1.2\div -0.8$ and $w_a\approx -1\div 1$. Note that when $w$ is not restricted to be constant the fit to the observations improves dramatically (see Huterer,  Cooray 2005). Indeed, the best fit is provided 
by models with $w<-1$ for $z<0.2$ ($1\sigma$ evidence); such a property would violate the null energy condition so that $p+\rho<0$, and the corresponding field is usually referred to as "Phantom energy" (see Alcaniz 2004; Carrol, Hoffman, Trodden  2003; Amendola 2004). Such a form of DE would most naturally lead to a Big Rip, i.e., to an infinite energy density increase in a finite time (Caldwell, Kamionkowski, Weinberg 2003). As for the  dynamical evolution, the fits to SNIa data mildly favour an evolving equation of state with $w_a>0$ (see, e.g., Feng, Wang, Zhang 2005; Upadhye, Ishak,  Steinhardt 2005), implying a crossing of the value 
$w=-1$ at redshift $\lesssim 1$ (this kind of evolving DDE was dubbed "Quintom", see, e.g.,  see Feng et al. 2005), a dynamics which differs from Quintessence and Phantom in determining the fate and the evolution of the Universe 
(see Guo et al. 2005) .

One important reason why the above observational constraints to DDE still allow large regions of the parameter space is they all involve either integrals of the Hubble expansion rate 
$H(a)=\dot a/ a$ over a large cosmic time (from $z\approx 1000$ to the present) or the value of $H(a)$ at relatively low redshifts $z\lesssim 1$. This would actually be sufficient to severely constrain DE if its equation of state was indeed constant, since its contribution to the energy density would become rapidly much smaller than the DM contribution (increasing like $a^{-3}$) at $z\approx 1$. However, time varying models with $w_a>0$ possess different behaviour from 
the $w=$ constant cases even at high redshifts  (see, e.g.,  Linder, Jenkins 2003)  both in the expansion rate $H(a)$ and in the growth factor $D=\delta(a)/ \delta(a_i)$ of DM overdensities $\delta\equiv \Delta \rho/\rho$ present at some initial 
time $a_i$. 

Extending the methods discussed above to higher redshifts to provide tighter constraints on DDE is a recognized priority for astrophysical research in the next future (see the National Academy of Science report 2010), and is involving a major effort of the international community. Tighter constraints on DDE equation of state will be provided by the NASA JDEM - WFIRST space mission (see Gehrels 2010) operating a 1.5-meter wide-field-of-view near-infrared-imaging and low-resolution-spectroscopy telescope; 
this will extend BAO and SNIa measurements to $z\approx 2$, yielding a reduction of a factor $\approx 5$ in the allowed area of the $w_0-w_a$ plane (see Frieman, Turner, Huterer 2008). The ESA space mission Euclid (see Laurejis et al. 2008) will 
consist of a 1.2m telescope placed in L2 orbit, and is optimised for weak gravitational lensing and BAO so as to achive measurements of the DDE equation of state with $\Delta w_a\leq 0.2$. A great statistical improvement will be achieved with the ground-based 8.4m  Large Synoptic Survey Telescope (LSST; Ivezic et al. 2010), that  will 
observe a 20 000 deg$^2$ region with a final limiting magnitude of $r\approx  27.5$, providing, e.g., an unprecedented sample of 
$\approx 3$ billion lensing galaxies. 

The above methods and experiments are all based either on the well-established standard Friedman-Laimaitre models
providing the dependence of distances on cosmological parameters, or on the evolution of linear density perturbations, for which we have a solid physical  description to be compared with observations. 
However, measuring  $w(a)$ over a wide range of cosmic times could in principle be achieved through observations involving 
high-contrast, non-linear cosmic structures. One example is the measurement of the redshift-space distortions, the imprint of large-scale peculiar velocities on observed galaxy maps (see Guzzo et al. 2008); since gravity-driven coherent motions are a direct consequence of the growth of structures, measurements of the redshift-space two-point correlation function can be mapped into estimates of the growth  factor $D(a)$. Alternatively, a measurement  of the DDE equation of state could be achieved 
by comparing the observed abundance of high-contrast cosmic structures, like clusters of galaxies with given DM mass (see Borgani \& Guzzo 2001), with the abundances computed assuming different background DDE (see Wang,  Steinhardt, 1998; Mota, Van Der Bruck 2004; Nunes, Mota 2006; Mota, Shaw, Silk 2008). The drawback of these methods is that the observed properties of such structures are related to their underlying DM distribution by the complex physics of gas and star formation; the uncertainties in the modelling of such processes are in general larger than the effects due to different assumed DDE cosmologies, and at present are not competitive with BAO and SNIa methods in constraining the DDE equation of state, although cluster abundances measured with the XMM  XXL survey (Pierre et al. 2011) and eROSITA  (Predhel et al. 2010; Cappelluti et al. 2010) are expected to provide substantial improvements in the next few years.  However, these methods probe the DE equation of state at $z\leq 1$. 

There are, however, objects that can be suitable for the above task, and that can probe DDE models at  higher redshifts. These are the most massive Black Holes (BHs) residing in the centre of massive high-redshift galaxies, and observed (Fan et al. 2006) to shine as Quasars (QSOs) already at early cosmic times ($z\approx 6$); these are the high-redshift, high-luminosity representatives of the Active Galactic Nuclei emitting over the whole electromagnetic spectrum due to the accretion of gas onto the Supermassive Black Holes (SMBHs), observed to be hosted in the bulges of local galaxies (see, e.g., Richstone et al. 1998). They are thought to originate from seed BHs of 
$\approx 100\,M_{\odot}$ produced by the collapse of PopIII stars at high redshift $z\gtrsim 20$ (see Madau \& Rees 2001), 
which subsequently grow through merging and accretion. Both such processes must be physically related with the growth of their host galaxies, since the BH masses and the properties of the host (e.g., with the stellar mass) are very tightly  correlated (Ferrarese \& Merrit 2000; Gebhardt et al. 2000). Besides, both processes are characterized by upper limits: 
the merging rate of BHs is limited by that of their host galaxies (see Volonteri, Hardt \& Madau 2003), 
which at high redshifts $z\gtrsim 6$ is smaller - but of the same order of magnitude - of the merging rate of the DM haloes collapsed from the peaks of the DM density field (see Lacey \& Cole 1993), primary related to the growth factor of DM perturbations;  the accretion rate is limited by the Eddington rate $\dot M_{Edd}=(10^{47}{\rm erg\,s^{-1}}/\epsilon\,10^{9}\,M_{\odot}\,c^2)\,M_{BH}$, 
where $\epsilon$ is the accretion efficiency and $M_{BH}$ is the BH mass; thus the maximal mass accretion rate 
is $\dot M=(1-\epsilon)\dot M_{Edd}$ yielding an exponential mass growth $M_{BH}(t)\propto exp[(1/\epsilon-1)\,t/t_{Edd}]$ 
where $t_{Edd}=4.5 10^7$ yrs.  

The discovery of BHs of masses  $\approx 3\,10^9\,M_{\odot}$ already in place at $z\approx 6$ (Barth et al. 2003; Willott, McLure, Jarvis 2003) clearly provides  
a challenge for cosmological models. Although such objects are extremely rare, with a number density $N\approx 10^{-9}\,Mpc^{-3}$, 
cosmological models must allow for the building up of such massive BHs from $10^2\,M_{\odot}$ seeds at $z\approx 20$, a huge mass growth in a short lapse of cosmic time (only 600 Myrs for the Concordance Cosmology). Our scope is to investigate which DDE models provide i) a cosmic time available for BH accretion since $z=6$ and ii) a growth factor (determining the BH merging rate) large enough to allow for the early building up of the observed population of  SMBHs. 

Using the abundance of  such objects at high redshifts $z\approx 6$ as a probe of DDE has several advantages: i) they are characterized by a {\it maximal} growth that can be easily modelled; ii) their maximal accretion  rate is {\it exponentially} sensitive to the cosmic time $t$;  iii) at high redshifts, the {\it maximal} merging rate of BHs is directly related to (and limited by) the growth factor of the galactic DM haloes where they reside;  iv) they form on the {\it exponential} tail of the DM mass distribution, so their predicted abundance is {\it exponentially} dependent on the mass grown by redshift $z\approx 6$, making the comparison with observations an extremely sensible probe. 

 To pursue our goal we shall develop an analytic approach to derive robust upper limits for the growth of the BHs mass functions up to $z\approx 6$ with different DDE background cosmologies. We shall compare the predicted abundance of 
BHs with masses $M_{BH}\geq 10^9\,M_{\odot}$ at such redshift with the observed number density, to derive exclusion
regions for the parameter space of DDE models which allow to reduce the regions allowed by other up-to-date observational probes (like SNIa, BAO, or CMB); in fact, we shall show that the abundance of high-redshift QSOs is able to 
disfavour some of the DDE models presently providing very good fits to the combined data from the above observations ($w<-1$ and $w_a\gtrsim 0.5$), and in particular the Quintom scenario. 

Our computation will provide solid constraints for DDE model provided three basic conditions are verified, at least for high BH masses $M_{BH}\geq 10^8\,M_{\odot}$ and for high redshifts $z\geq 6$. These are: 1) the BHs form from primordial seeds with mass $\approx 10^2\,M_{\odot}$ originating from the collapse of PopIII stars at $z\geq 20$; 2) the Eddington limit for the accretion rate holds at all times; 3) the average accretion efficiency verifies $\epsilon\geq 0.1$. We shall discuss the validity of such conditions in Sect. 2 and in the Conclusions.

The plan of the paper is as follows: in Sect. 2 we describe our analytical approach to derive upper limits on the 
growth of the BH mass function; in Sect. 3 we shall compare the predictions for different  DDE models with the
observed abundances of $z\approx 6$  SMBHs, and derive severe exclusion regions for the DDE parameters; Sect. 4
is devoted to discussion and conclusion. 

\section{The Evolution of the Black Hole Mass Function}

Here we present an analytical approach to follow the evolution of the BH mass function in a cosmological framework. 
In such a framework, galaxies form from the collapse of the peaks of primordial DM density field, and grow through repeated merging events.  At high redshifts, the large densities of baryons collapsing in the DM haloes allow for rapid gas cooling (Rees, Ostriker 1977), so that copious cold gas reservoirs are available for both star formation and BH accretion (see, e.g., Baugh 2006 for a review). 
Subsequently, BHs grow by accretion of cold galactic gas and by merging with other BHs following the coalescence of the host galaxies. 
We describe below how we model the corresponding evolution of the BH mass function, maximizing the 
parameters related to their accretion, for any background cosmology. 

\subsection{Model Set Up}

Cosmic structures form in DM haloes collapsed from tiny overdensities ($\delta\rho/\rho\approx 10^{-5}$) of a primordial Gaussian density field (see Padmanabhan 1993; Peebles 1993, Peacock 1999, Coles and Lucchin 2002). Measurements of the CMB first with COBE (Smoot 1992) and later with BOOMERANG (De Bernardis et al. 2000), MAXIMA (Hanany et al. 2000), WMAP (Hinshaw et al. 2003) provided increasing evidence for such a scenario, and allowed for a detailed measurement of the  power spectrum of the density field; complementary high-precision measurements of the power spectrum were obtained through the galaxy correlation function (Percival et al. 2006; Tegmark et al. 2006; Pope et al. 2004, Cole et al. 2005, Padmanabhan et al. 2006). 
The resulting spectrum is characterized by a rms density fluctuation $\sigma (M)$ inversely dependent on the filtering mass scale $M$, so that  small-scale peaks  collapse earlier (on average) as they reach the threshold $\delta_c$ for non-linear evolution
(see Peebles 1993); as larger and larger regions of the density field become non-linear, the previously formed haloes merge to form progressively larger haloes. The probability ${d^{2}P(M\rightarrow M')/dM^{'}dt} $ that a DM haloes with mass $M$ is included into a larger halo $M'$ in a time step $dt$ is provided by the Extended Press \& Shechter formalism (see Bond et al. 1991;  Bower 1991; Lacey \& Cole 1993) 
\begin{equation}
{d^2 P(M\rightarrow M',t)\over d 
M'\,dt} = \Bigg[{\sigma^2\over\sigma'^2(\sigma^2-
\sigma'^2)}\Bigg]^{3/2}
{\delta_c\,e^{-{\delta_c^2(t)\,(\sigma^2-\sigma'^2)\over 2\,D^2(t)\,\sigma^2\,\sigma'^2}} \over \sqrt{2\pi}\,D^2(t) }
\,
\Big|{d\sigma'^2\over dM'}\Big|\Big|{dD(t)\over dt}\Big|~, 
\end{equation}
where $\sigma$ and $\sigma'$ are the variance of the density fluctuations corresponding to 
the masses $M$ and $M'$, respectively. Note that the merging rate depends critically on the {\it growth factor} 
$D(t)$ of DM density perturbations in the linear regime, which in turn is sensible to the DDE background cosmology; its expression in a flat Universe with a cosmological constant is 
given in Carroll, Press, Turner (1992) as a function of $\Omega$. The critical threshold $\delta_c$ for the collapse
of density fluctuations also depends weakly on cosmology, ranging from $\delta_c=1.65$ for the Concordance Cosmology 
to $\delta_c=1.68$ of a critical $\Omega=1$ cosmology (see Mainini, Maccio, Bonometto, Klypin 2003). 
 
When two DM galactic haloes merge, the dynamical friction against the DM background and the stars, and the viscous effects from the surrounding gas drag the hosted BHs towards the centre of the most massive DM halo, where they form a binary system, loose orbital energy through the emission of gravitational waves and eventually coalesce (Volonteri, Hardt, Madau 2003; Callegari et al. 2009; see Colpi, Dotti 2009 for a review).  Several processes may delay (or even inhibit) the coalescence of BHs inside merged DM haloes (Milosavlejevic \& Merritt 2001), so that the DM merging rate provides an effective {\it upper limit} to the BH merging.

Besides merging, SMBHs grow through accretion of cold galactic gas. The relation between the dynamical evolution of galaxies in the DM haloes and the fraction of gas feeding the BH accretion depends on the complex physics linking the 
formation of accretion disks on sub-pc scales to the processes involving the evolution of the galactic gas. 
However, an  {\it upper limit} to the accretion luminosity is provided by the Eddington value $L_{Edd}=4\,\pi\,G\,M\,c/k$ 
(where $k$ is the electron scattering opacity) yielding for the accretion rate a maximal value $\dot M_{Edd}=L_{Edd}/\epsilon\,c^2=(10^{47}{\rm erg\,s^{-1}}/\epsilon\,10^{9}\,M_{\odot}\,c^2)\,M_{BH}$, where $\epsilon$ is the accretion efficiency, specified by the fractional binding energy of the innermost stable circular orbit about the hole. The latter in turn 
ranges from $6\,r_g$ to $r_g$ (where $r_g=G\,M_{BH}/c^2$ is the gravitational radius), for non-spinning and for maximally rotating holes, respectively; thus, in principle, the accretion efficiency may range from $\epsilon\approx 0.06$ (for non-spinning holes) to $\epsilon\approx 0.45$ for maximally rotating holes. 
The SMBHs maximal accretion rate is thus given by 
\begin{equation}\label{medd}
$$\dot M_{BH}=(1-\epsilon)\,\dot M_{Edd}={(1-\epsilon)\over \epsilon}\,{M_{BH}\over t_{Edd}}$$ 
\end{equation}
where  $t_{Edd}=4.5\,10^7$ yrs (the complementary accreted mass-energy $\epsilon\,\dot M_{Edd}$ being emitted as radiation). 

We can describe the joint effect of merging and accretion onto the growth of SMBHs in terms of their conditional mass function $N(M_{BH},M,t)$, i.e., the number density of BHs with mass in the range $M_{BH}\div M_{BH}+\Delta M_{BH}$ hosted in DM haloes with mass in the range $M\div M+\Delta M$ at cosmic time $t$. The evolution in a time step $\Delta t$ of such quantity is given by 

\begin{eqnarray} \label{eq}
N(M_{BH},M,t+\Delta t)&=&p(M,t) \times N(M_{BH}-(\dot M_{BH}\Delta t),M,t)+(1-p(M,t))\times N(M_{BH},M,t)\nonumber\\\
&+&\Delta t \int_{M_{BH,inf}}^{M_{BH}} \int_{M_{inf}}^{M}N(M_{BH}^{'},M^{'},t)\, \frac{d^{2}P(M^{'}\rightarrow M)}{dM^{'}dt}  \frac{N(M_{BH}-M_{BH}^{'},M-M^{'},t)}{N_{H}(M-M^{'},t)}dM_{BH}^{'} dM^{'} \\
&-&\Delta t \times  N_{H}(M,t)\int_{M}^{M_{sup}} \frac{d^{2}P(M\rightarrow M^{'})}{dM^{'}dt} dM^{'} \nonumber\
\end{eqnarray}
Here $N_H(M,t)= \sqrt{2/ \pi} ({\rho\,\delta_c / M^2\,\sigma})\,
exp[- {  \delta_c^2(t) / 2\,D^2(t)\,\sigma^2}   ]\,|{dln \sigma/ dln M}|$ is the number density of virialized structures of mass $M$ at the cosmic time $t$  given by the  Press \& Shechter (1974) formula, depending on the variance of the power spectrum $\sigma (M)$ , on the growth factor $D(t)$, and on the  mean matter density of the Universe $\rho$. 

The first term on the r.h.s. of eq. (\ref{eq}) accounts for the statistical effect of accretion: the function $p(M,t)$ denotes the probability that a BH hosted in a halo of mass $M$ is in an active accretion phase at time $t$: in such a case, its BH mass
$M_{BH}$ is increased by an amount $\dot M_{BH}\,\Delta t$.  Of course, maximizing the accretion corresponds to assuming $p(M,t)=1$ and an Eddington accretion rate $\dot M_{BH}$ given by eq. (2).
The second term on the r.h.s. of eq. (\ref{eq})  represents the construction of BHs due to merging. 
The increment in the number of BHs 
with mass $M_{BH}$ in haloes with mass $M$ is given by the number of BHs with smaller mass $M'_{BH}$ in haloes with  
smaller DM mass $M'$ weighted with the
halo merging probability $d^2P(M\rightarrow M')/dM'dt)\,dM'\,dt$  and with the probability $N(M_{BH}-M_{BH}^{'},M-M^{'},t)/N_{H}(M-M^{'},t)$ of finding a  BH with complementary mass $M_{BH}-M'_{BH}$ in the second merging halo. 
The third term on the r.h.s. of eq. (\ref{eq})  represents the desctruction of BHs of the considered mass $M_{BH}$ due to the coalescence of its host halo with mass $M'$ to form a larger DM halo.
Of course, we are again maximizing the BH merging rate because we are assuming that BH merging immediately follows the coalescence of DM haloes. Nevertheless, we expect our upper limit for the merging rate to provide a fair description of the actual  rate, since at high redshifts the dynamical friction is very efficient in dragging the BHs in merging haloes toward the centre of the merger (see Bough 2006 for a review);  this is a specific property of hierarchical scenarios, characterized by a two-phase growth of cosmic structures, with major merging at high redshifts $z\gtrsim 3-4$ rapidly followed by coalescence of substructures, and accretion of small clumps at low $z\gtrsim 2$, as noted by several authors (e.g., Zhao, Jing \& B\"orner  2003; Diemand, Kuhlen \&  Madau 2007; Hoﬀman et al. 2007; Ascasibar \& Gottloeber 2008). Thus, although eq. (\ref{eq})  will yield an upper bound to the evolution of the mass function, we
expect its solutions to be close to the actual mass function at high redshifts $z\gtrsim 6$. 

We now proceed to describe the strategy we adopt for solving eq. (\ref{eq})  for different background DDE cosmologies.  

\subsection{Strategy and Solutions} 

We aim at solving eq. (\ref{eq})  with a choice of the initial conditions and of the accretion rate and probability that maximize the 
growth of SMBHs for a given background cosmology; this will allow us to exclude all DDE models which do not allow for the building up of the observed abundance of massive $M\approx 3\,10^9\,M_{\odot}$ BHs at $z=6$. We focus on the highest redshift for which observed abundances are available, since this enhances the constraints on cosmological models (their maximal accretion rate is exponentially  sensitive to the cosmic time, see eq. (\ref{medd}) ) and because at such redshifts, the {\it maximal} merging rate of SMBHs is directly related to (and limited by) the growth factor of the galactic DM haloes where they reside. 

Before performing an extensive study of the effects of DDE models on the evolution of the mass function, we discuss here several critical points: the initial conditions, the role of the BH spin (determining the value of the radiation efficiency $\epsilon$ entering the accretion rate in eq. (2)), and the accretion probability ($p(M,t)$ in eq. (\ref{eq}) ) related to the duty cycle of the accretion phases of the BH. Our aim is not to provide a detailed description of the above points, but rather to point out realistic upper limits for the growth of SMBHs set by the above processes. 

\subsubsection{Initial conditions} \label{ic} 
According to Madau \& Rees (2001), seed BHs may have formed from the very first generation of stars. 
These would collapse from a metal-free gas leading to a top-heavy IMF, as suggested by several authors (Bromm, Coppi, Larson 1999, 2002; Abel et al. 2000; Yoshida et al. 2006; Gao et al. 2007) corresponding to very massive stars (VMS) with masses larger than 100 $M_{\odot}$. The stars that form with masses smaller than $\approx 300$ would end up into pair-instability SNae, and their stellar cores would be entirely disrupted leaving no remnants (Kudritzki, Puls, 2000; Fryer et al. 2001). Higher mass stars instead collapse leaving    remnant BHs with mass $M_{BH}\gtrsim 150\,M_{\odot}$. Accroding to Madau \& Rees (2001), the primordial generation of stars could form at redshifts $z\gtrsim 20$ in DM haloes with virial temperatures $\approx 10^3$ K where gas can cool due to collisional excitation of $H_2$. 

 To set our initial condition we proceed following Madau \& Rees (2001), Volonteri, Hardt \& Madau (2003). We start populating at $z=20$ the peaks of the primordial density perturbation field, starting with the higher mass, rarer peaks. We populate progressively lower mass  haloes and we compute the corresponding amount of photons produced in the Lyman-Werner (LW) band (11.2-13.6 eV) by the progenitor zero-metallicity stars; these photons easily dissociate $H_2$ molecules, so populating the DM haloes with VMS results in a UV background that suppresses molecular cooling throughout  the Universe (Omukai \& Nishi 1999; Haiman, Rees, Loeb 1997), thus inhibiting the formation of a much more numerous population of VMS and hence of seed BHs; further cooling will take 
place later due to atomic processes in haloes with virial temperature $T_{vir}$ $\gtrsim$ 10$^4$ K corresponding to DM halo masses $M\geq 10^{7.9}$[(1+z)/20]$^{-3/2}$ $M_\odot$. The fraction of baryons in VMS stars 
is $f=300\,M_{\odot}\,F_{DM}/M\,\Omega_b$, where $F_{DM}$ is the fraction of mass in the DM density field collapsed in DM haloes populated with VMS (and hence with seed BHs), and $M$ is the minimum mass for such haloes. The VMS in such haloes produce $\gtrsim 10^5$ LW photons during a lifetime of $2\,10^6$ yrs, so that the number of LW photons per baryons will be $3\,10^7\,M_{\odot}\,F_{DM}/M\,\Omega_b$. The minimum mass $M$ corresponds, for a Gaussian DM density field, to a fraction of populated haloes $F_{DM}=erfc\big[\delta_c/\sqrt{2}\,D(t)\sigma(M)\big]$.  Thus requiring the LW flux not to exceed $10^{-2}$ photons/baryon (again, we choose the value in the range $10^{-2}\div 10^{-4}$ provided by Haiman, Abel, Rees (2000) maximizing the abundance of BHs) yields  for the lower mass $M$ of the DM haloes populated with BHs a value $M\geq M_{min}\approx 10^7\,M_{\odot}$ for a $\Lambda$CMD cosmology (entering the growth factor $D(t)$ in $F_{DM}$), corresponding to populating the peaks above 3$\sigma$. This corresponds to a cosmic density of VMS ( and hence seed BHs) at $z=20$  {\it larger} by about a factor of 2.5 than that usually adopted ( $\sim$1 VMS $ Mpc^{-3}$, e.g. Volonteri, Hardt, Madau 2003), so again we are  {\it maximizing} the mass that BHs may later acquire from merging of the seeds.
 
  Similar BHs densities are obtained when applying the same procedure at redshifts higher than 20, corresponding to populating lower mass haloes; a physical limit is set by the existence of a minimum redshift-dependent mass ($\approx 10^6\,M_{\odot}$ at $z=20$) below which the $H_2$ production rate is not sufficient to cool the gas efficiently 
 (Fuller \& Couchman 2000). 

The above procedure sets the initial conditions for eq. (\ref{eq})  by determining the initial density $N(M_{BH},M,t_{in})$; this is taken to assume the Press \& Schechter value when $M_{BH}=100\,M_{\odot}$ and $M\geq M_{min}$, and to be zero elsewhere. 
The above procedure for determining the initial conditions is repeated for all the different DDE background cosmologies we will consider.  

We do not consider in this paper the alternative scenario envisaging the collapse of  SMBHs with mass $\approx 10^5\,M_{\odot}$ directly out of dense gas (see Begelman et al. 2006; Lodato \& Natarajan 2007; see also Volonteri \& Rees 2005); this scenario would require avoiding fragmentation, i.e., high UV flux to avoid cooling and low-metallicities, conditions that may be satisfied in rare cases (Dijkstra et al. 2008; Omukai 2001; Bromm \& Loeb 2003); the metal-free condition only would be 
 incompatible with the presence of nearby luminous galaxies (Omukai 2008). Even if fragmentation was avoided, the cold gas flows inward at low velocities, near the sound speed of a few km $s^{-1}$ , with a correspondingly low accretion rate of 
∼ 0.01 $M_{\odot}$/yr. This results in conditions nearly identical to those in the cores of lower-mass 
minihalos; extensive ultra–high resolution simulations have concluded that the gas then forms a single 
$\approx 150 \,M_{\odot}$ star (Abel et al. 2002; Bromm et al. 2002; Yoshida et al. 2008) 
or perhaps a massive binary (Turk et al. 2009).  A path toward the solution of the above models could rely on additional heating by a primordial magnetic field (Sethi et al. 2010). 

\subsubsection{The role of BH spins} 
The BH angular momentum $J$ is generally described in terms of the spin parameter $\hat a\equiv J\,c/G\,M_{BH}^2$ ranging from 
$\hat a=0$ for non-rotating BH to $\hat a=1$ for maximally rotating BHs. It enters 
the radiation efficiency $\epsilon$ in the Eddington accretion rate (eq. 2), which increases in the range $0.06-0.42$ with  
$\hat a$ increasing from 0 to 1.  As recalled in Sec. 2.1, larger values of $\epsilon$ suppress the Eddington accretion rate.

Observationally, a wide range of values is allowed for $\epsilon$, which is usually measured through the Soltan argument (Soltan 1982) or from synthesis models (e.g. Merloni \& Heinz 2008) relating the evolution of the AGN luminosity functions (i.e., the statistical
evolution of the BH accretion rates) to the local mass distribution of relic BH masses resulting from the accretion history. These measurements yield $\epsilon\approx 0.06\div 0.16$ (see, e.g., Elvis et al. 2002; Marconi et al. 2004; Wang et al. 2006; Treister \& Urry 2006; Hopkins, Richard \& Henquist 2007; Merloni \& Heinz 2008; Yu, Lu 2008; Shankar et al. 2008), corresponding to  $\hat a= 0.05-0.9$. However, the lower limits on $\epsilon$ are derived by the above authors assuming merging to be negligible in the mass growth of BHs. While computations show that most of the final BH mass results from accretion (see below), merging is crucial in splitting the accretion history of a BH into many progenitors which can simultaneously accrete at the Eddington limit. Taking into account the enhancement in the accretion power due to  such effect, the recent observational limits given by Merloni \& Heinz (2008) can be recast in the form $\epsilon\geq 0.08/\xi_0(1-\xi_m)$, where
$\xi_0$ is the present BH mass density normalized to the value $4.3\,10^5\,M_{\odot}\,Mpc^{-3}$, and
$\xi_m$ is the fraction of final BH mass density contributed by splitting the BH accretion history into many progenitors due to merging. In the framework of cosmological hierarchical growth of DM haloes considered here, $90 \%$ of the massive haloes $M\geq 10^{12}M_{\odot}$ had at least one major merging event at $z\geq 6$ (De Lucia et al. 2004; Stewart et al. 2008; see also Lacey \& Cole 1993; Gottl\"ober, Klypin, Kravtsov 2001), so that $\xi\gtrsim 0.5$ yielding values of $\epsilon$ above $0.1$ even assuming the largest value of $\xi_0=1.5$ allowed by observations. Of course, this conclusion holds only when observations are interpreted in the framework of hierarchically growing DM haloes, which constitutes one of the assumptions of our computations.
On the other hand, theoretical arguments have long been favouring large values of the spin parameter $\hat a$, close to the maximum value $\hat a=1$; in fact, even if $\hat a\approx 0$ initially, accretion can spin the BH up to $\hat a\approx 1$ (see Volonteri  2006) since the BH aligns its spin with the angular momentum of the disk due to a combination of the Lense-Thirring effect with viscous dissipation (Bardeeen \& Petterson 1975; Scheuer \& Feiler 1996; Natarajan \& Pringle 1998). However, Moderski \& Sikora (1996), King \& Pringle (2006 see also King, Pringle, Hofmann 2008)  have noted that spin-down can occur when counter rotating material is accreted, a condition that can occur if the angular momentum of the disk is strongly misaligned with that of the BH and the accretion disc actually fragments due to self-gravity, breaking the accretion into a series of small uncorrelated episodes ("chaotic accretion"),
each feeding the BH with a mass $m_{acc}=\hat a\,M_{BH}\,(r_s/r_w)^{1/2}\approx 10^{-4\div -3}\,M_{BH}$ (here $r_s$ is the Shwarzschild radius, and $r_w$ is the distance of the warp produced by the Bardeen-Petterson process in the accretion disk, Bardeen \& Petterson 1975).
In such a case, the Lens-Thirring effect can actually counter-align the BH and the accretion disk angular momenta; although co-rotating and counter-rotating accretion are equally probable, chaotic accretion tends to produce a decrease of the spin since the innermost stable orbit for a retrograde orbit is at larger radii than for a prograde orbit, and the transfer of angular momentum is more efficient in the former case. BH-BH merging also affects the spin distribution; a wide exploration of the merging parameters has shown that distributions with average $\langle \hat a\rangle \approx 0.7$ (corresponding to $\epsilon=0.1$) results from merging of BHs with different masses and initial angular momenta (Berti \& Volonteri 2008), a value close to that holding for equal mass, non-spinning BHs. The combined effect of all the above processes has been investigated in recent works, where the effect of the environment during galaxy merging is also taken into account. Simulations by Dotti et al. (2006; 2007, see also Volonteri 2010) address the evolution of spin in gas-rich mergers, our assumed trigger for BH accretion, following in detail the evolution of the BH angular momentum. In gas-rich mergers, the gas settles into a dense circumnuclear disk in which the BHs relative orbit becomes
circular and corotating; the BHs then align their spins with the angular momentum of their orbit and hence with that of the large scale gaseous disc (Liu 2004; Bogdanovic, Reynolds \& Miller 2007; Dotti et al. 2010). As a result BHs acquire spins $\hat a\geq 0.7$ (Berti \& Volonteri 2008; Kesden, Sperhake \& Berti 2010); magneto-hydrodynamic simulations specifically aimed at deriving the evolution of BH spin at high redshifts support large values of $\hat a\geq 0.8$ (Shapiro et al. 2005).
Independent works (Fanidakis et al. 2010) including the evolution of BH angular momentum into a semi-analytic model of galaxy formation have shown that, even assuming gas-poor mergers, the high-mass ($M_{BH}\geq 3\,10^8\,M_{\odot}$) BHs end up with large values of $\hat a\geq 0.7$ due to the effect of BH merging, with chaotic accretion effectively acting on smaller BH undergoing minor accretion episodes at lower redshift (e.g., on the BHs harboured in local spiral galaxies), as also suggested by Sikora et al. (2007).
The above works indicate that if the BH accretion is connected to the hierarchical merging histories of DM haloes,
then $\epsilon \geq 0.1$ (corresponding $\langle \hat a\rangle=0.7$), at least for the large BH masses we are interested in. Since our model is defined in such a framework, we shall take $\epsilon=0.1$, the lower value in the range where theoretical expectations overlap with observational uncertainties.
Larger spin values would result in a slower evolution of the BH mass function, and hence in stronger exclusion constraints for DDE models.

\subsubsection{The duty cycle} 
Such a quantity enters the probability of accretion $p(M,t)$ in eq. (3) . If merging triggers the BH accretion, this can be computed as the probability that a halo of mass $M$ had a major merging within 
the duration of the last accretion episode. Assuming that the accretion lasts through a time $\tau$ , the probability for a halo with mass $M$ to merge with any  DM halo with mass larger than $qM$ can be calculated using eq. (2.22) of Lacey \& Cole (1993); for  $q=0.01$ (equivalent to considering all the mergers contributing to 99\% of the DM mass $M$) and estimating $\tau\approx r/v\approx 3.8 \times 10^7$ yrs (at $z=10$, the ratio depending only on $z$, see 
Baugh et al. 2010) as the galaxy crossing time (as a typical duration of encounters), would yield probability ranging from 1 for DM haloes more massive than $10^{14}\,M_{\odot}$ (hosting the most massive BHs) down to 0.6 for haloes with mass $M=10^6\,M_{\odot}$. However, as in the previous case, we aim at maximizing the growth of BHs to obtain robust exclusion regions for DDE models, so that we will use $p(M,t)=1$ in eq. (\ref{eq})  for all $M$ and $t$. The solution obtained with this choice will be close to those obtained using the above estimates of $p(M,t)$ for large BH masses $M_{BH}\geq 5\,10^8\,M_{\odot}$, so that our exclusion regions for DDE models are not affected by uncertainties in the estimate of the duty cycle. 
%However, the faint end of the BH mass function at $z=6$ will be affected by the choice of $\tau$; although such a region of the BH mass distribution is not probed by present observations, the comparison of our predicted mass function with future measurements of the slope of the BH mass distribution at the faint end will provide a constraints for the BH duty cycle at high redshifts $z\approx 6$ (see Sect. 3.2).

\subsection{The cosmological models}
We assume a spatially flat, homogeneous and isotropic universe filled by non-relativistic matter plus a dark energy component. The cosmological expansion is described by the Friedmann equation:
%\begin{equation}\label{friedmann}
%H^{2}=\left(\frac{\dot{a}}{a}\right)^2=H_0^2\left( \sum_i \Omega_{0i} \exp \left(-3\int_{a0}^{a}\frac{1+w_i(a^{'})}{a^{'}}da^{'}\right)+\left(\frac{a_0}{a}\right)^2(1-\sum_i \Omega_{0i})\right)
%\end{equation}
\begin{equation}\label{friedmann}
H^{2}=H_{0}^{2}[\Omega_{M}a^{-3}+\Omega_{\Lambda}e^{3\int_{a}^{1} (1+w(a^{'}))da^{'}/a^{'})} ]
\end{equation}
where H$_0$=100$h$ km/s/Mpc is the present value of the Hubble expansion factor,  $\Omega_{M}$ is the dimensionless matter density at the present epoch, $\Omega_{\Lambda}$=1-$\Omega_{M}$  denotes the DE density parameter, and $w(a)$ is the equation of state parameter of the DE.

%,  $\Omega_{0i}=\rho_{0i}/\rho_{0c}$ is the current density parameter of the i-th component of the universe, $\rho_{0c}$ =3H$_0^2$/8$\pi$G is the critical density at present time, w$_i$ is the equation of state parameter of the i-th component. The first term in eq. (\ref{friedmann}) includes densities associated to each constituent of the universe while the second term accounts for any possible deviation from flat geometry. We consider flat cosmological models, so the curvature term will be equal to zero. \\
%The lookback time from the present day (t$_0$) to an object at redshift z$_1$ is given by:
%\begin{equation}\label{htz}
%t_0-t_1=\int_{1/(1+z_1)}^{1} \frac{da}{aH(a)}.
%\end{equation}
%the age of the universe is obtained by taking the z$_1$ $\rightarrow$ $\infty$ (t$_1$ $\rightarrow$0) limit.
The growth of the cosmological matter perturbations in the linear regime for sub-Hubble scales (and for standard gravity) is governed by the equation:
\begin{equation}\label{growth}
\ddot{\delta}+2H\dot{\delta}-4\pi G\rho_m \delta=0
\end{equation}
where $\delta$=$\delta \rho$/$\rho$ is the matter density contrast, and the dot denotes derivative with respect to time.  The perturbations grow according to a source term involving the amount of matter able to cluster and are restricted by a friction term (Hubble drag) arising from the expansion of the universe.   In general, it is preferred to express the linear solution of eq. (\ref{growth}) as a function of the linear growth factor $D$, defined as  the ratio  of the perturbation amplitude at some scale factor relative to some initial scale factor: $D$=$\delta$(a)/$\delta$(a$_i$) . 

The solutions to eqs. (4) and  (5) depend on the equation of state through the parameter $w$ entering eq. 
(\ref{friedmann}). For our analysis, we considered general DE models, in which the DE equation of state may vary with time. We use the Chevallier-Polarski-Linder parametrization (Chevallier \& Polarski 2001, Linder 2003) to describe the evolution in terms of the scale factor $a$: 
\begin{equation}\label{w}
w(a)=w_0+w_a(1-a)=w_0+w_a\,{z\over 1+z}
\end{equation}
where the parameter $w_0$ represents the value of $w$ at the present epoch: $w_0$=$w(a=1)$, while $w_a$ corresponds to its look-back time variation: $w_a$=-$dw/da$; In the above parametrization, the standard $\Lambda$CDM cosmology corresponds to $w_0$=-1 and $w_a$=0. Kowalski et al. (2008), using data from SNe, CMB and BAO put constraints on these parameters corresponding to $w_0\approx -1\pm$0.2 and $w_a\approx \pm$1. 

Using this parametrization eq. (\ref{friedmann}) becomes:
\begin{equation}\label{}
H^{2}=H_{0}^{2}[\Omega_{M}a^{-3}+\Omega_{\Lambda}a^{-3(1+w_0+w_a)}e^{3w_a(a-1)} ].
\end {equation}
The above equation yields the cosmic time $t$ corresponding to a given redshift $z$ in any DE model:
\begin{equation}\label{htz}
t(z)=\int_{z}^{\infty} \frac{dz'}{H(z')(1+z')}.
\end{equation}

As for the growth factor, the solution to eq. (\ref{growth}) for the $\Lambda$CDM case is given by Carroll, Press, Turner (1992) in the form :
\begin{equation}
\delta(a)=\frac{5 \Omega_M}{2a}\frac{da}{d\tau}\int_0^a \left(\frac{da^{'}}{d\tau}\right)^{-3}da^{'}
\end{equation}
where $\tau$=H$_0$t;. For the DE models we use the parametrization to the solution given in Linder (2005):
\begin{equation}\label{d_par}
\frac{\delta(a)}{a}=exp\left(\int_{0}^{a}[\Omega_M(a)^{\gamma}-1]dlna\right) 
\end{equation}
where  $\Omega_M(a)=\Omega_M a^{-3}/(H(a)/H_{0})^2$, and $\gamma$  is the growth index, given by the fitting formula  (Linder 2005):
\begin{eqnarray} \label{gamma}
\gamma=0.55+0.05(1+w(z=1)) & w(z=1)\geq 1   \nonumber\ \\ 
\gamma=0.55+0.02(1+w(z=1)) & w(z=1)< 1  \,.
\end{eqnarray}
This parametrization reproduces the growth behaviour to within 0.1\%-0.5\% accuracy for a wide variety of dark energy cosmologies (Linder 2005, Linder \& Cahn 2007)  and allows for a rapid scanning of the parameter space of DDE models. We have checked that the constraints we derive on $w_0$-$w_a$ with such a parametrization are indistinguishable from those obtained by integrating eq. (5). 
We  normalize the growth factors of the DE models to their high redshift behaviour ($D$=$\delta$(a)/$\delta$(a=0)),  corresponding to WMAP normalization of the matter power spectrum. 
This procedure yields different values of $\sigma_8$, the power spectrum normalization in terms of the variance of the density field smoothed over regions of 8 $h^{-1}$ Mpc, for different DE  models.
For the $\Lambda$CDM cosmology this corresponds to $\sigma_8=0.8$ for the local ($z=0$) variance of the density field  (Larson et al. 2011). 
 
Both the time-redshift relation (eq. (8)) and the growth factors (eq. (10)) corresponding DE models deviate mildly from the cosmological constant case when the equation of state is negatively evolving  with redshift $w_{a}$ $<$0, while models with  $w_{a}$ $>$0 yield growth factors and cosmic times lower than those predicted in the  $\Lambda$CDM case. This is because in the $w_{a}$ $>$0 models  the influence of DE at early times is strong even at high redshift (see eq. (7)), yielding shorter ages (eq. (8)) and implying a delay in the growth of DM perturbations compared  the $\Lambda$CDM case with important implications on the  structure formation; both effects are illustrated in Fig. 1. Note that, since at high redhshifts $w=w_0+w_a$ in this parametrization, any model with $w_0+w_a\leq 0$ will not yield a matter dominated early Universe, altering the sound horizon in conflict with the observations (see Kowalski et al. 2008).

We use the above relations (eqs. (8), (9) (10)) in the computation of the number density of BHs at high redshifts after eq. (3);  
we adopt the following set of cosmological parameters from the WMAP 7-year mission results (Larson et al. 2011): $\Omega_M$=0.27, 
 $\Omega_{\Lambda}$=0.73; $H_0=71\,$km$s^{-1}$Mpc; the power spectrum determining $\sigma (M)$ has 
 spectral index $n$=1.

\begin{figure}
\begin{center}
\includegraphics[width=8 cm]{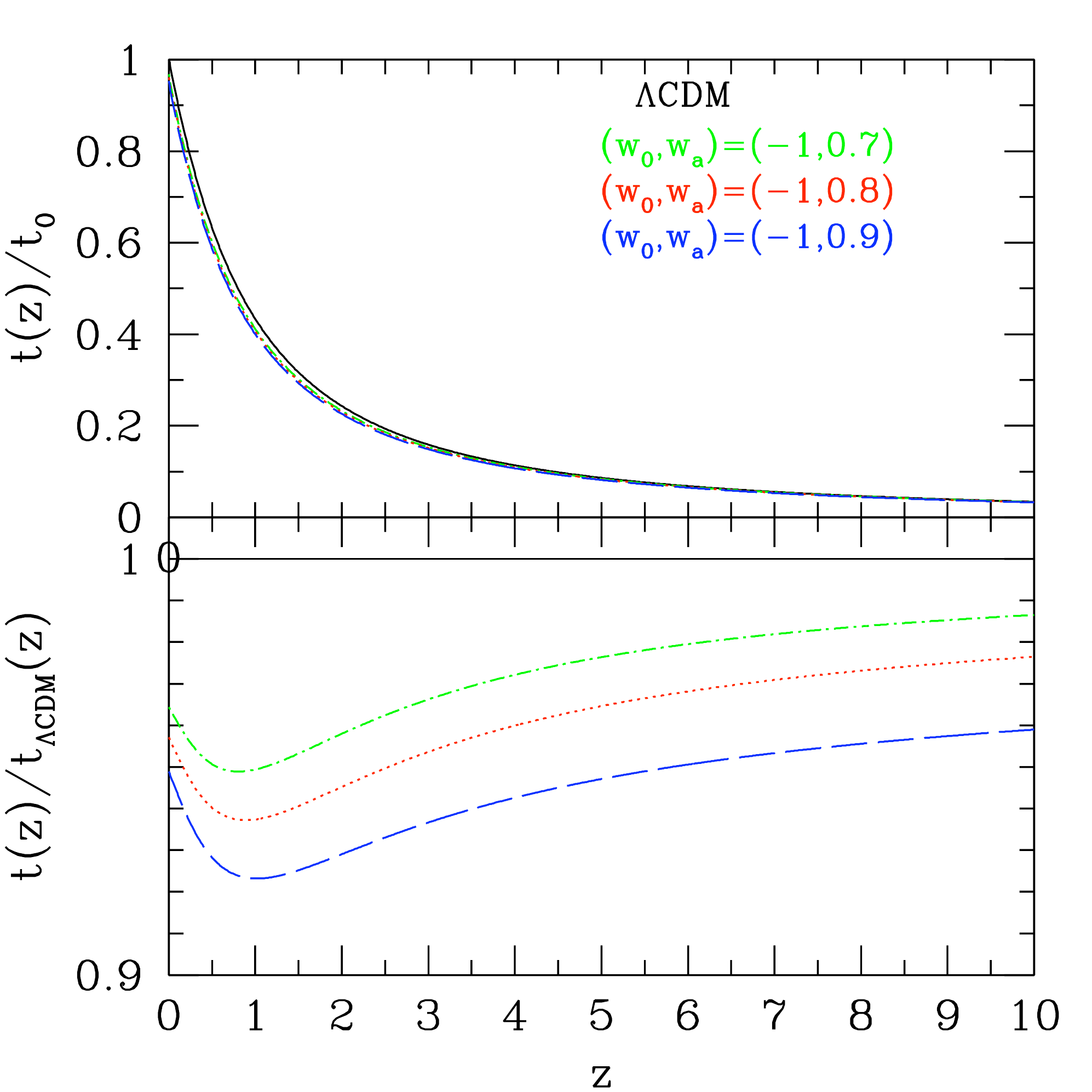}
\includegraphics[width=8 cm]{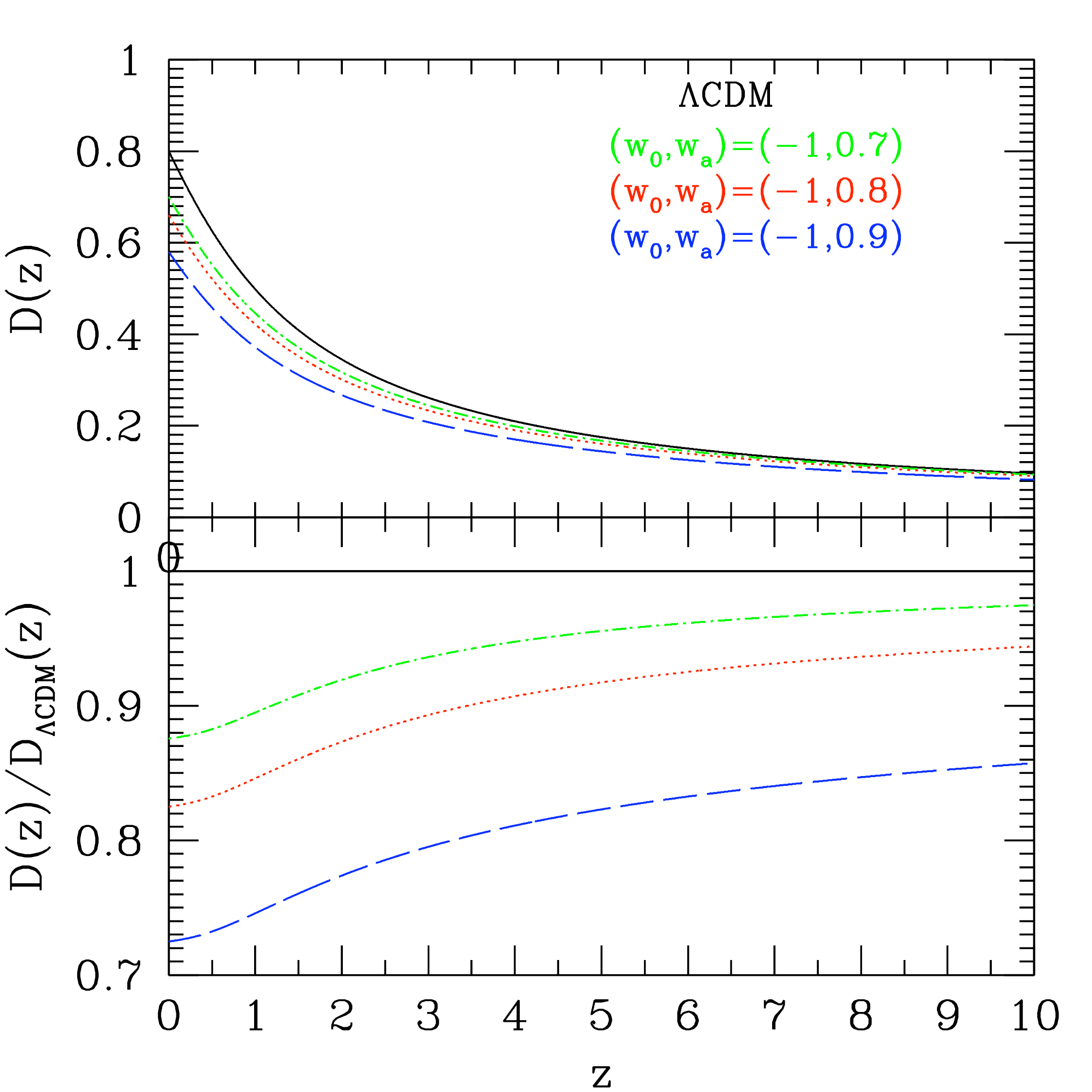}
 \caption{The redshift evolution of the cosmic time (left) and growth factor (right), predicted by the DDE models with  ($w_{0}$,$w_{a}$)=(-1,0.7)  (dot-dashed line) , ($w_{0}$,$w_{a}$)=(-1,0.8)  (dotted line), and ($w_{0}$,$w_{a}$)=(-1,0.9)  (long-dashed line)  are compared with the $\Lambda$CDM case
($w=1$ solid line) in the top panels; the lower panels show the same quantities normalized to the $\Lambda$CDM values.}
\end{center}
\end{figure}

\section{Results}

\subsection{The evolution of the BH mass function in dark energy cosmologies}

In order to compute the evolution of the number density $N(M_{BH},M,t)$ of BHs with mass $M_{BH}$ hosted in haloes with mass $M$ we  integrated  numerically eq. (\ref{eq}). The time step was set to $\Delta$t=3.3 10$^{-4}$ H$_0^{-1}$ and we adopted a 
logarithmic grid for black hole and halo masses spanning 8 orders of magnitude from $log M_{BH}/M_{\odot}=2$ and $log M/M_{\odot}=6$
with a grid size $\Delta$log$M$=0.04. We derived the BH mass function at cosmic time $t$ by integrating $N(M_{BH},M,t)$ over the host halo mass $M$. Before solving the full eq. (3) with realistic initial conditions,  we first tested our code against analytic solutions that are available for simple cases. The case $p(M,t)$=0 corresponds to ignoring the BH mass accretion; starting with 
BH masses equal to a constant fraction $f$ of the halo mass $M$ in each DM halo, the BH number density at time $t$ 
should be given by the Press \& Schechter expression for masses $fM$, since in this case the only mode for BH growth is the merging of the host haloes (and of the hosted BHs) preserving the halo to BH mass ratio, and the halo mass distribution at any time is given 
Press \& Schechter form recalled in Sect. 2.1. We have checked that in this case our numerical solution of eq. (3) remains close to the analytic  Press \& Schechter form over the entire BH mass and redshift ranges explored in this work. We have also checked that 
in the opposite limit of no BH merging and constant BH accretion rate $r$, the numerical solutions of eq. (3) at time $t$ correspond to a shift of the initial mass distribution by an amount $M+r\,t$.

Next, we proceeded to compute the complete evolution of $N(M_{BH},M,t)$ due to both BH merging and accretion; the initial conditions are set as described in detail in Sec. 2.2.1.The BH mass function $\Phi(M_{BH},t)$,  i.e. the number density of BHs with mass in the range $M_{BH}\div M_{BH}+\Delta M_{BH}$ at cosmic time $t$, is computed by integrating $N(M_{BH},M,t)$ over the halo mass.
Since our goal is to determine the maximal abundance of BHs allowed by a given cosmology, we assumed that BH merging immediately follows the coalescence of DM haloes (Sect. 2.1), and that BHs undergo a continuous growth ($p(M,t)$=1, see sect. 2.2.3) through gas accretion at the Eddington limit (eq. (2)). Conservatively, the only physical limit to BH accretion we consider is the presence of dense, cool gas 
at the centre of galaxies. This is relevant at high redshifts $z\gtrsim 20$, when the gas in the DM halos can cool due to collisional excitation of $H_2$ in DM haloes with virial temperatures $T_{vir}$ $\approx 10^3$ K. In these circumstances, we set up our initial conditions as discussed in Sect. \ref{ic}. At lower redshifts, the primordial generation of stars formed from this cold gas emits UV photons that suppress further molecular cooling throughout  the Universe (as discussed in detail Sect. \ref{ic}); thus, after $z\sim$20, further cooling (and hence accretion given by eq. 2) is possible only within more massive haloes with virial temperatures 10$^4$ K $\lesssim$ $T_{vir}$ $\lesssim$ 10$^6$ K, where the gas cools via the radiative decay of collisionally excited atoms, and in haloes with  higher virial temperatures where the dominant  cooling mechanism is the emission of Bremmsstrahlung radiation (see e.g. Baugh 2006). We also checked that at all times the mass accreted by BHs remains smaller than the cold gas mass contained in the host DM halos. The latter is calculated by assuming  an isothermal halo gas density profile and integrating  out to the cooling radius, defined as the radius where the cooling time of the gas is equal to the age of the universe (White \& Frenk 1991).

The full evolution of the BH mass function for the case of constant $w=-1$ (cosmological constant) is shown for $z\leq 10$ in the left panel of Fig. 2. The relative importance of accretion and merging in the growth of BHs is illustrated in the right panel, where we compare $\Phi$  at $z$ $\simeq$ 6 (solid line), with the mass functions obtained by assuming that between $z$=10 and $z$=6 the BH growth is
only due to merging ($\Phi_{mer}$ dotted line), or to accretion ($\Phi_{acc}$ dashed line). The evolution of the mass function due  to  continuous Eddington-limited accretion is a strict translation of the mass function towards higher BH mass, while the effect of BH merging is to flatten the slope of the mass function, as it contributes to make high mass BHs from many smaller progenitors. Although the difference between $\Phi_{mer}$ and $\Phi_{acc}$ is huge (the maximum BH mass in the two cases differs by about three orders of magnitude), this does not mean that merging  is not important in the formation of massive BHs, as it can be inferred by comparing the solid and dashed lines in Fig. 2.  Indeed, the number density of BHs obtained with accretion plus merging  is greater than that obtained without merging by a factor $\sim$ 20  
for $M_{BH}$ $\simeq$ 10$^9$ $M_{\odot}$, and by a factor $\sim$ 100 for $M_{BH}$ $\simeq$ 3 10$^9$ $M{\odot}$. This is because  merging enhances the effective accretion power by distributing the BH accretion in the progenitors of massive BHs; 
thus, simultaneous Eddington-limit accretion can occur in BHs that later will assemble into a unique, larger object.

\begin{figure}
\begin{center}
\includegraphics[width=8 cm]{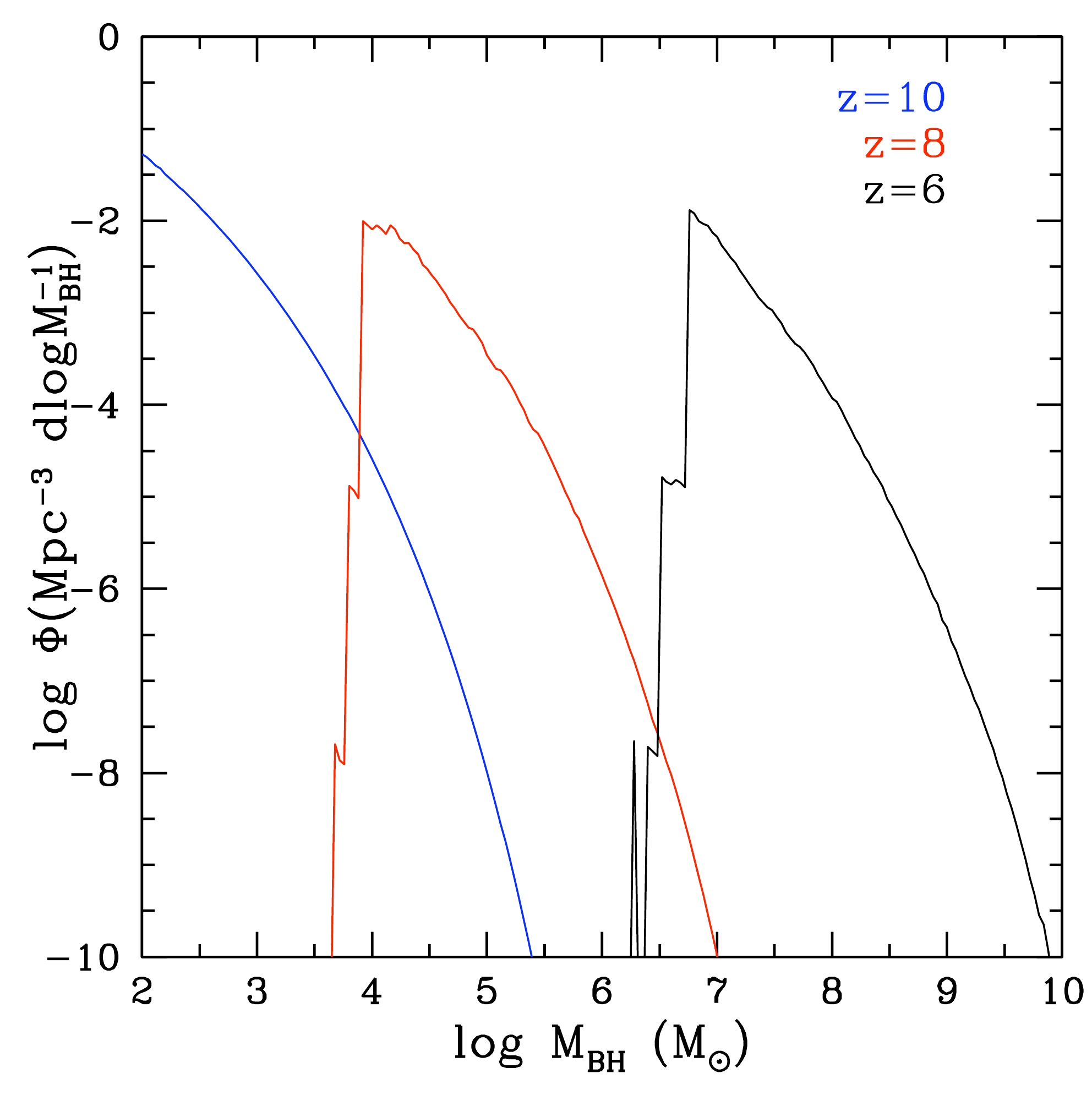}
\includegraphics[width=8 cm]{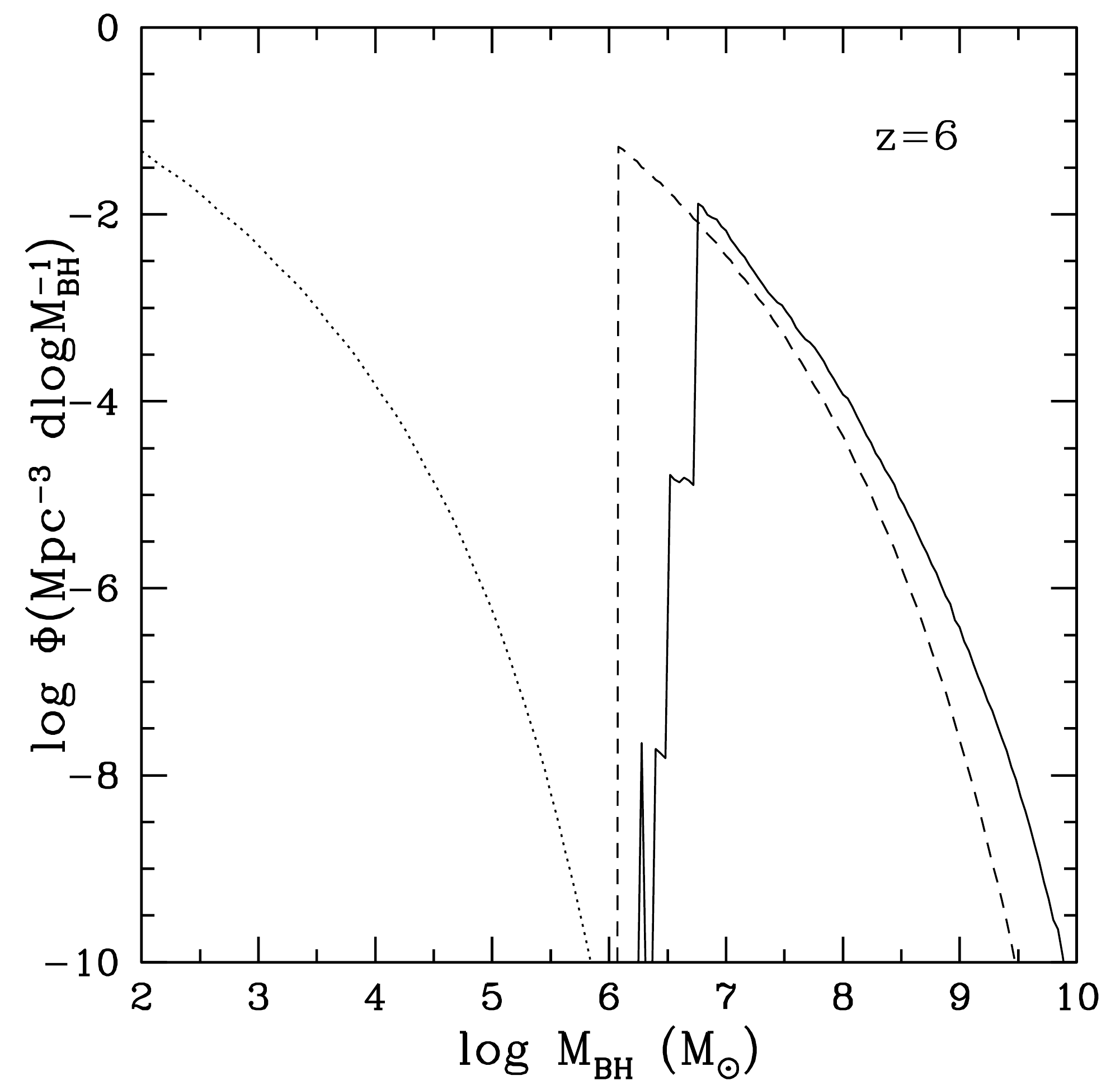}
\caption{Left panel:  The BH mass function at $z$=10,8 and 6 predicted by the $\Lambda$CDM cosmology. Right panel: The BH mass function at $z$=6 obtained assuming BH growth due to only merging (dotted line), BH growth due to only accretion (dashed line), and BH growth due to merging plus accretion (solid line). }\label{BHMF_evol}
\end{center} 
\end{figure}

The role of different DE cosmologies in the growth of BHs is illustrated in Fig. 3, where we show the comparison between the maximal abundance of massive BHs  at z=6 allowed by the $\Lambda$CDM cosmology (solid line), with that allowed by three different DDE cosmologies corresponding to  ($w_{0}$, $w_{a}$)=(-1, 0.9)  (dashed line),  ($w_{0}$, $w_{a}$)=(-1, 0.8)  (dotted line), ($w_{0}$, $w_{a}$)=(-1, 0.7)  (dot-dashed line). 
In this figure, we only compare the $\Lambda$CDM case with DDE cosmologies with $w_a>0$. In fact, as 
discussed in Sect. 2.3, DDE  models with  $w_{a}$ $<$0  yield linear growth factors and ages very close to the cosmological constant model at $z$=6 and therefore they predict the same BH number density at this redshift. In contrast, DE models with $w_{a}$ $>$0 predict lower values of $D$($z$=6) and $t$($z$=6) (see Fig. 1); although the differences with respect to the $\Lambda$CDM case are at most a factor of 0.9 in cosmic age and a factor of 0.7 in growth factor, the exponential sensitivity of the BH mass function to such quantities (discussed in Sect. 2.1) results in appreciably different BH number densities. In particular, for increasing values of $w_a$, the larger DE densities at high redshift entering the expansion rate (the second factor on  the r.h.s. of in eq. 7) result in slower growth factors and short ages at $z=6$ (see Fig. 1): the first mechanism results into a steepening of the mass function, as it reduces the BH merging rate, while the second mechanism reduces the time available for  BH accretion and results in a shift of the mass function toward lower masses. 

Note that in principle also the critical threshold for the collapse of density fluctuations, $\delta_c$ depends weakly on cosmology. Several authors (e.g. Mainini, Maccio, Bonometto, Klypin 2003, Pace, Waizmann, Bartelmann 2010 ) derived $\delta_c$ for different DE scenarios, indicating that models with $w_a>0$ have values of  $\delta_c$ which are at most of $\sim$1,2\% greater that the value assumed in  $\Lambda$CDM  cosmology ($\delta_c$=1.65).  Since the larger the value of the critical density $\delta_c$,  the steeper is the associated mass function, assuming a larger $\delta_c$ would imply smaller number densities of massive objects.
In order to maximize the predicted BH abundance, we conservatively assume $\delta_c$=1.65 for all models.
%However the  BH number density obtained with  the largest allowed value of  $\delta_c$ is similar to that obtained with $\delta_c$=1.65, %thus we did not calculate $\delta_c$ for each DDE models but we 

\begin{figure}
\begin{center}
\includegraphics[width=10 cm]{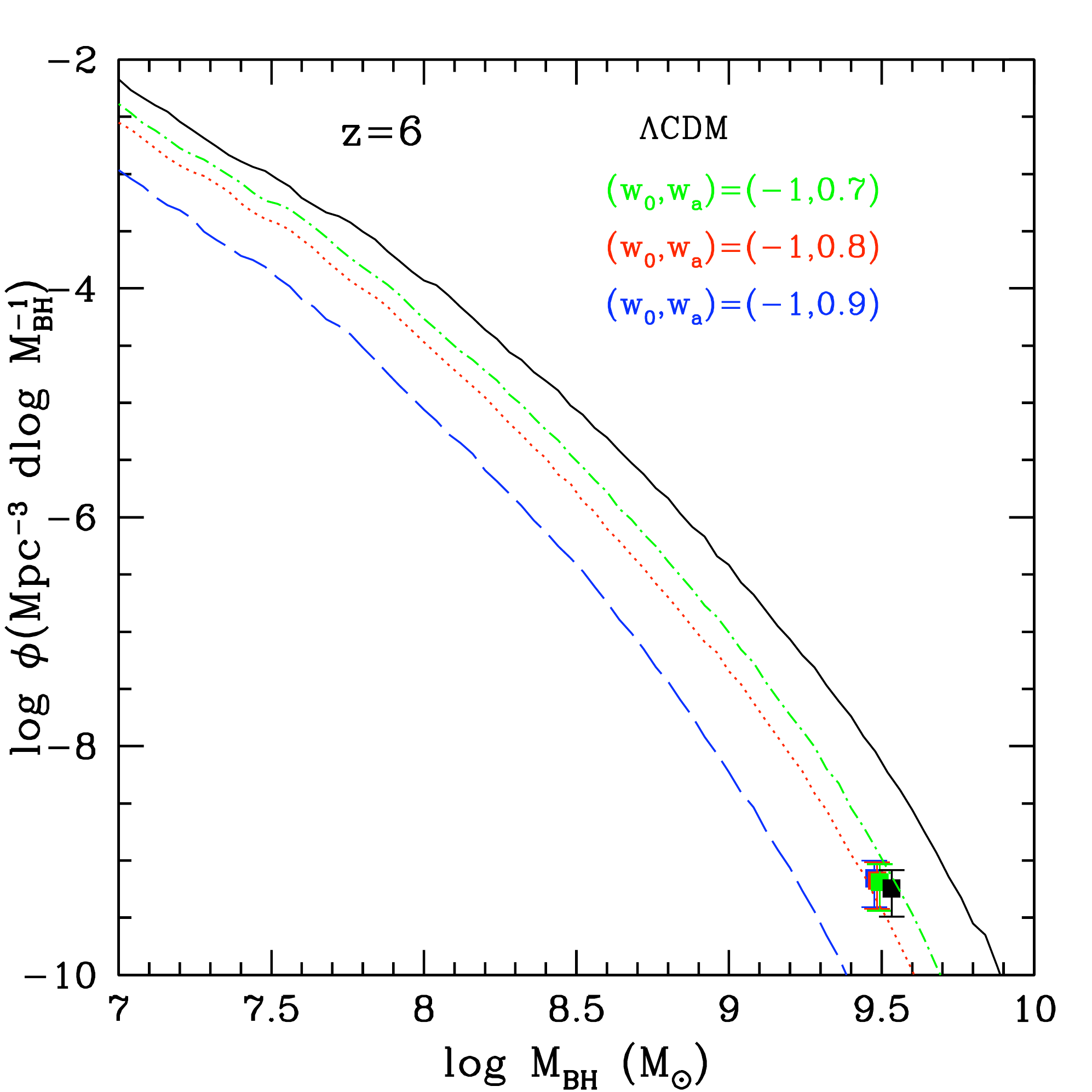}
\caption{The SMBH mass function at $z$=6 predicted by the $\Lambda$CDM cosmology (solid line), and by three different DDE cosmologies corresponding to  ($w_{0}$,$w_{a}$)=(-1,0.7)  (dot-dashed line) , ($w_{0}$,$w_{a}$)=(-1,0.8)  (dotted line), and ($w_{0}$,$w_{a}$)=(-1,0.9)  (long-dashed line) is compared with the observed number densities (squares) derived from the SDSS QSO luminosity functions (see text). 
In order to compare the predicted and observed number densities, we re-scaled the latter to each cosmology as it is indicated by the coloured squares in Fig. 3. The shift of the observational point toward lower masses and higher densities is due to the lower luminosity distances and comoving volumes predicted by the cosmological models with respect the $\Lambda$CDM case.  }\label{BHMF_cosmo}
\end{center}
\end{figure}

The squares in Fig. 3 represents the observed BH number density derived from the $z$ $\sim$6  Sloan Digital Sky Survey (SDSS) QSO luminosity function from 
Jiang et al. (2009). We consider only the brightest luminosity bin, corresponding to the more massive objects, since this enhances the
  constraints on the cosmological models. Willott et al. (2010a) also derived the $z$ $\sim$6 QSO luminosity function by combining the Canada-France high-$z$ Quasar Survey (CFHQS) with the SDSS samples. However, in both cases the bright end of the luminosity 
function is sampled by the SDSS main sample described in Fan et al. (2000, 2001, 2003, 2004, 2006). As discussed in Willott et al (2010a), the differences in the bright end are due to different luminosity bin sizes  and different SDSS sample  (Jiang et al. included some unpublished data) used by the authors.
Deriving the BH number density from the observed luminosity function  is not a trivial procedure as it requires several assumptions on the
Eddington ratio, bolometric corrections, corrections for inactive BHs and corrections for obscured quasars.  However,  in order to meet our goal of excluding DDE models that provide  too slow evolution of the BH number density, it is sufficient to derive lower limits on the observed abundance of BHs. Thus,  we assume a BH duty
cycle of 1, and we estimate the mass of the central black holes by assuming that they are emitting at the Eddington luminosity, as indicated by the Eddington ratio distribution obtained by Willott et al. (2010b). 
 To estimate the Eddington BH masses of these high-$z$ QSO we use the bolometric luminosities obtained by Jiang et al. (2006) by combining multiwavelength  observations (from radio to X-ray)  available in the literature (for details on such a procedure see, e.g., 
Fiore et al. 2011 ).
%We use the bolometric correction from  et al. (2006) to calculate the bolometric luminosity from the monochromatic emission at 1450 $\AA$; this is consistent with 
%that derived by Jiang et al. (2006) who combined multiwavelength  observations to determine the spectral energy distribution of 13 $z$ $\sim$6 SDSS. Although  the uncertainty on the bolometric correction is about a factor 2, 
The reliability of the above BH mass estimates is supported by the comparison with independent measurements of the virial BH mass. These have been obtained  from the Mg II line width and the continuum luminosity at  3000 $\AA$ of  two of the brightest sources in the SDSS main sample  (SDSS J114816.64+525150.3 and SDSS J083643.85+005453.3) by Willott et al. (2003) and Kurk et al. (2007). Their estimates of virial BH 
masses $M_{BH}$ $\simeq$ 3$\times$ 10$^9$ $M_{\odot}$ are very close to our derivation. 
The presence of dust in the high-$z$ quasar environments (e.g. Maiolino 2004, Jiang et al. 2006, Gallerani et al. 2010)  constitutes another source of uncertainty in the estimate of the BH mass from the AGN luminosity. Dust could lead to underestimate the QSO number densities, since heavily obscured AGNs could be undetected by optical surveys.  In detected sources, the observed luminosity could be reduced by dust exinction thus biasing the Eddington and virial BH mass estimates to lower values. Conservatively,  we do not correct the observed luminosity function for the aforementioned effects, correcting for these effects would only make our constraints on the cosmological models stronger.

To derive exclusion regions for the DDE parameters $w_0$ and $w_a$ we first solved eq. (3) for a  grid of cosmological DDE parameters,  then we computed the corresponding BH number density at $z=6$, and compared it with the observed abundance, thereby extending the comparison illustrated in Fig. 3 to the range  $-1.5\leq w_0\leq -0.5$ and $-2.5\leq w_a\leq 1.5$. 
Since  in our approach we maximized the computed growth of SMBHs (eq. (3))  for a given background cosmology, and we adopted conservative values for the observed abundances, we can firmly exclude all the cosmological models that predict number densities lower than the observed ones. In Fig. 4 we show the resulting exclusion regions in the $w_0$-$w_a$ plane; 
the three contours (the upper diagonal stripes) correspond to $w_0$-$w_a$ combinations that produce BH number densities  
{\it below} the observed value by a factor 1, 1.5, and $>$2 $\sigma$ (from bottom to top). Interestingly, our results provide constraints complementary to those obtained by  combining Supernovae, CMB  and BAO. These are shown by the 
ellipses representing the 68.3\%, 95.4,\% and 99.7\% confidence levels on  $w_0$-$w_a$ obtained by Kowalski et al. (2008, left panel), and by the more recent analysis by Amanullah et al. (2010, right panel). While the latter leave the time evolution of the state parameter $w_a$ almost unconstrained, the BH abundance provide much stronger constraints on $w_a$ than on the present value $w_0$. 
Combined with the existing constraints, our results significantly restrict the allowed region in DDE parameter space, strongly disfavouring 
DDE models that do not provide cosmic time and fast growth factor (Fig. 1) large enough to allow for the build-up of the observed abundance of BHs; in particular, models with $-1.2\leq w_0\leq -1$ and positive redshift evolution $w_a\gtrsim 0.8$ - consistent with previous constraints (and indeed mildly favoured according to some recent analysis, see  Feng, Wang, Zhang 2005; Upadhye, Ishak,  Steinhardt 2005;  and the analysis of the "Gold" dataset in Perivolaropoulos 2006) - result instead disfavoured by our independent constraints from BH abundance. Such range of parameters corresponds to "Quintom" DDE models (see Introduction), with $w$ crossing $-1$ starting from larger values.  For models with $w_0\gtrsim -1$, our results exclude DDE with an equation of state rapidly evolving with $z$, so that $-dw/da=w_a\geq 0.8$.
This limit has an impact on a wide class of models with a "freezing" behaviour of the DE scalar field (see Caldwell, Linder 2005; Linder 2006; see also the Summary). SUGRA inspired models (Brax \& Martin 1999), which are well fitted by $w_0\approx -0.82$ and $w_a\approx 0.58$ 
(Linder 2003) are disfavoured owing to the combination of our constraints with previous observational constraints (Fig. 4).

\begin{figure}
\begin{center}
\includegraphics[width=8 cm]{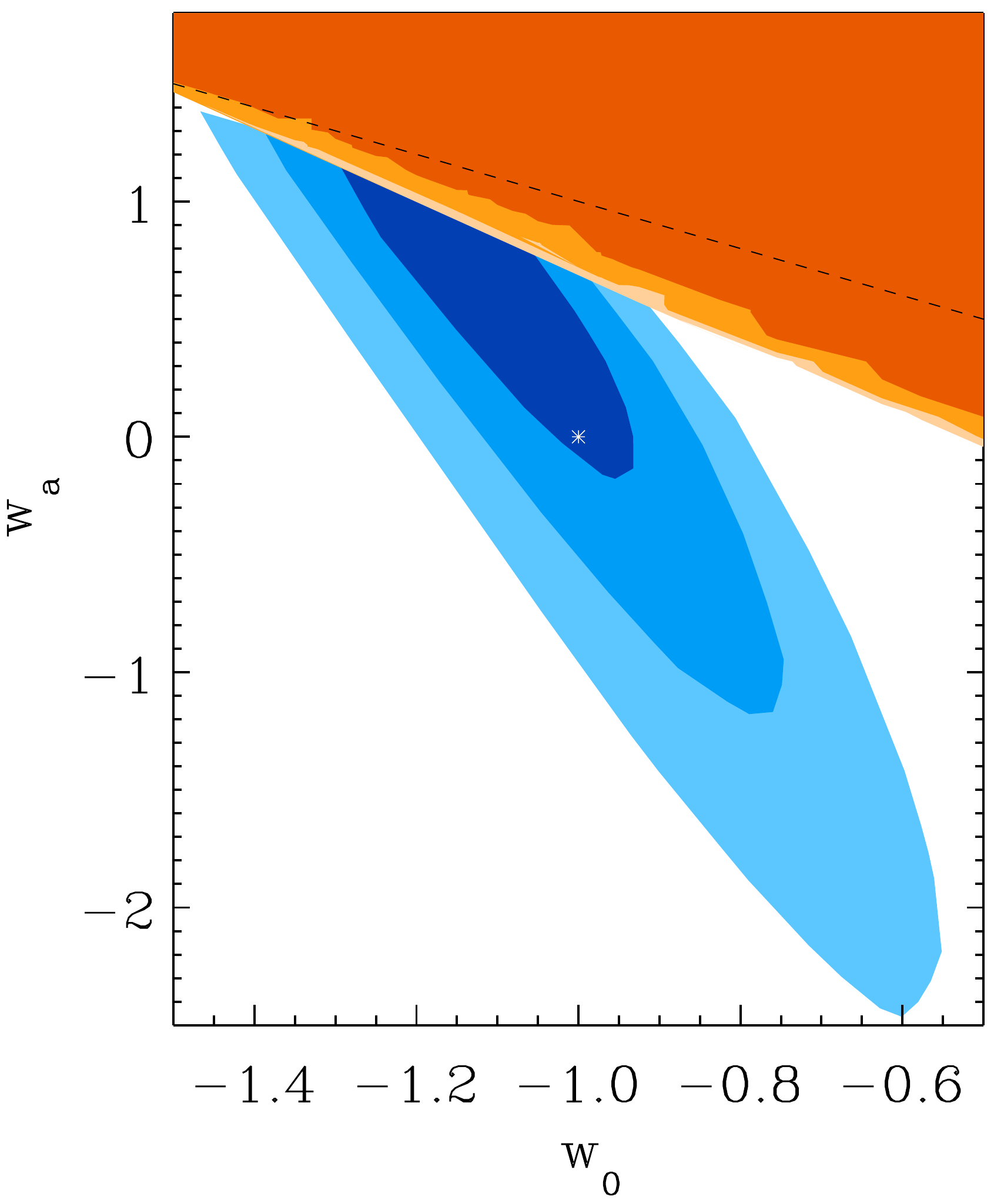}
\includegraphics[width=8 cm]{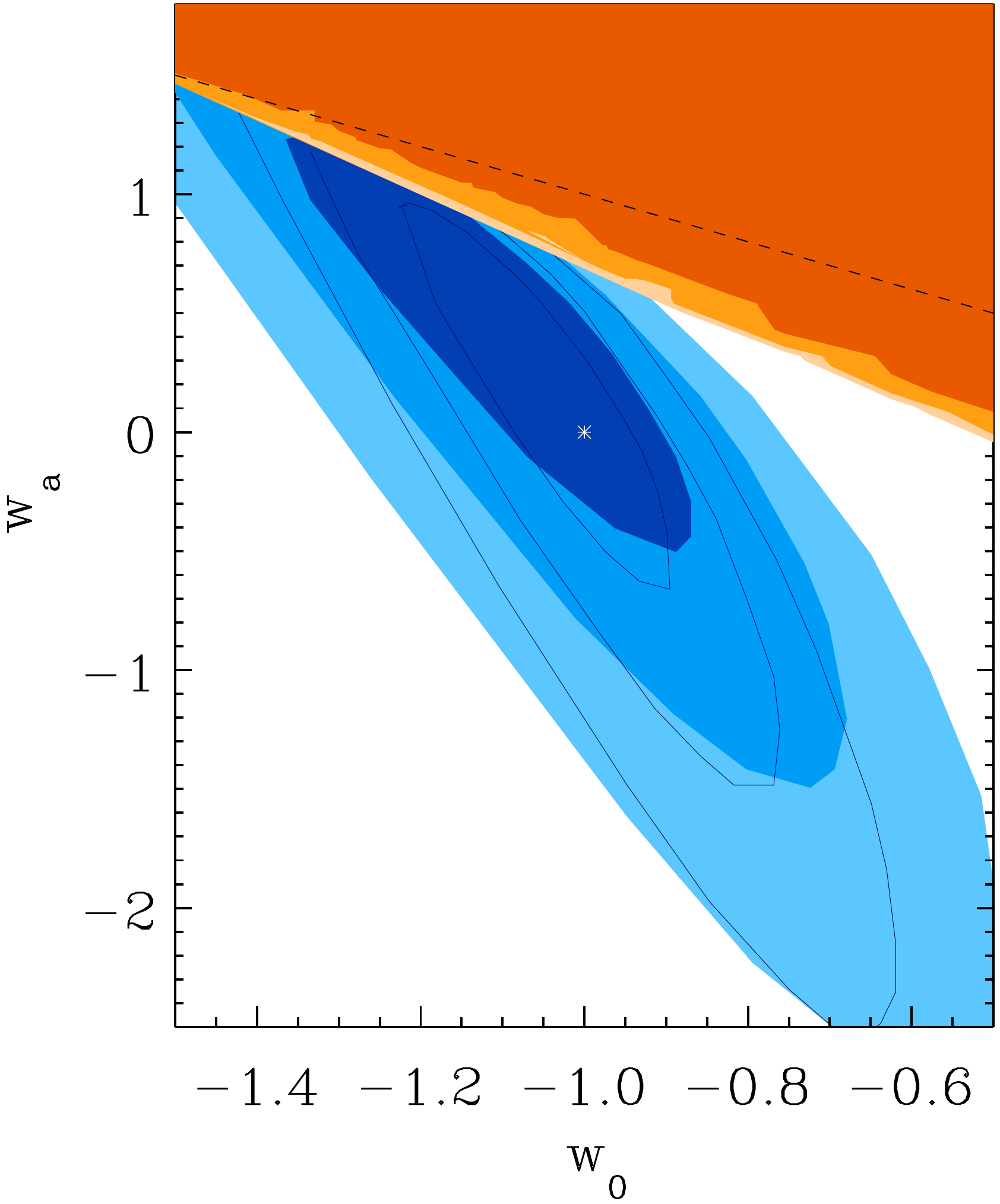}
\caption{The exclusion regions in the $w_0$-$w_a$ plane derived by our analysis (upper shaded region) with initial conditions and 
parameters set as described in Sect. 2.2. The filled contours correspond to the the 1, 1.5, and $>$2 $\sigma$ exclusion (from bottom to top); an useful analytic formula for the (1-$\sigma$) exclusion region is  $w_a\geq -3/2\,w_0-3/4 $
 The shaded ellipses represent the 68.3\%, 95.4,\% and 99.7\% confidence level on $w_0$-$w_a$ obtained by combining Supernovae, CMB  and BAO constraints, using the "Union"  
(Kowalski et al. (2008, left panel) and "Union 2" (Amanullah et al. 2010, right panel) compilation of SNae Ia data. In the right panel the solid contours represent the 68.3\%, 95.4,\% and 99.7\% confidence level on $w_0$-$w_a$ when systematic errors are not included. 
Values of $w_a$ above the dashed line $w_0+w_a=0$ correspond to violation of early matter  domination (see Sect. 2).}
\end{center}
\end{figure}

\subsection{Testing the reliability of the method, and comparison with semi-analytic results}

In this section we address how uncertainties in the assumed BH and halo properties affect the constraints derived for the DDE cosmological models. First, we study how these constraints change in response to different choices of the BH seed parameters. Indeed, the abundance of SMBH at $z$=6 depends on the abundance of seed BHs imposed at some higher redshift.  As described in Sect. 2.2.1 we adopted a scenario in which seed BHs are the end-product of very massive stars (VMS),  formed by primordial gas due to molecular hydrogen cooling. In this context,  the BH density  is related to  the density of the primordial stars  formed in the early haloes.  We calculated the VMS number density by requiring that the number of Lyman-Werner (LW) photons per baryon emitted by  primordial stars is sufficient  to prevent H$_2$ cooling (and hence further star formation) throughout the universe. This condition on the LW photon flux determines a minimum mass for the DM halo hosting the seed BH.  Therefore, in our approach, the parameters describing the BH seed population are: i) the seed BH mass (M$_{seed}$), ii) the redshift at which BH seeds are formed ($z_{in}$) , and iii)  the minimum mass of the DM halo hosting the seed BH (M$_{inf}$) ; iv) the mass distribution of the DM haloes hosting the seeds at the initial time.
 
The seed black hole mass depends on the mass of the VMS. Theoretical models of primordial star formation (McKee \& Tan 2008, see also Bromm et al. 2009 and reference therein) which take into account the  possible feedback processes that operate during the formation of the protostar predict a final stellar  mass  in the range 60-300   M$_{\odot}$   for reasonable values of the entropy and angular momentum of the pre-stellar gas. This implies a population of primordial BHs with masses M$_{seed}$ $\simeq$ M$_{VMS}$/2 $\lesssim$150 M$_{\odot}$ as the expected end-product of  pregalactic star formation.

\begin{figure}
\begin{center}
\includegraphics[width=5 cm]{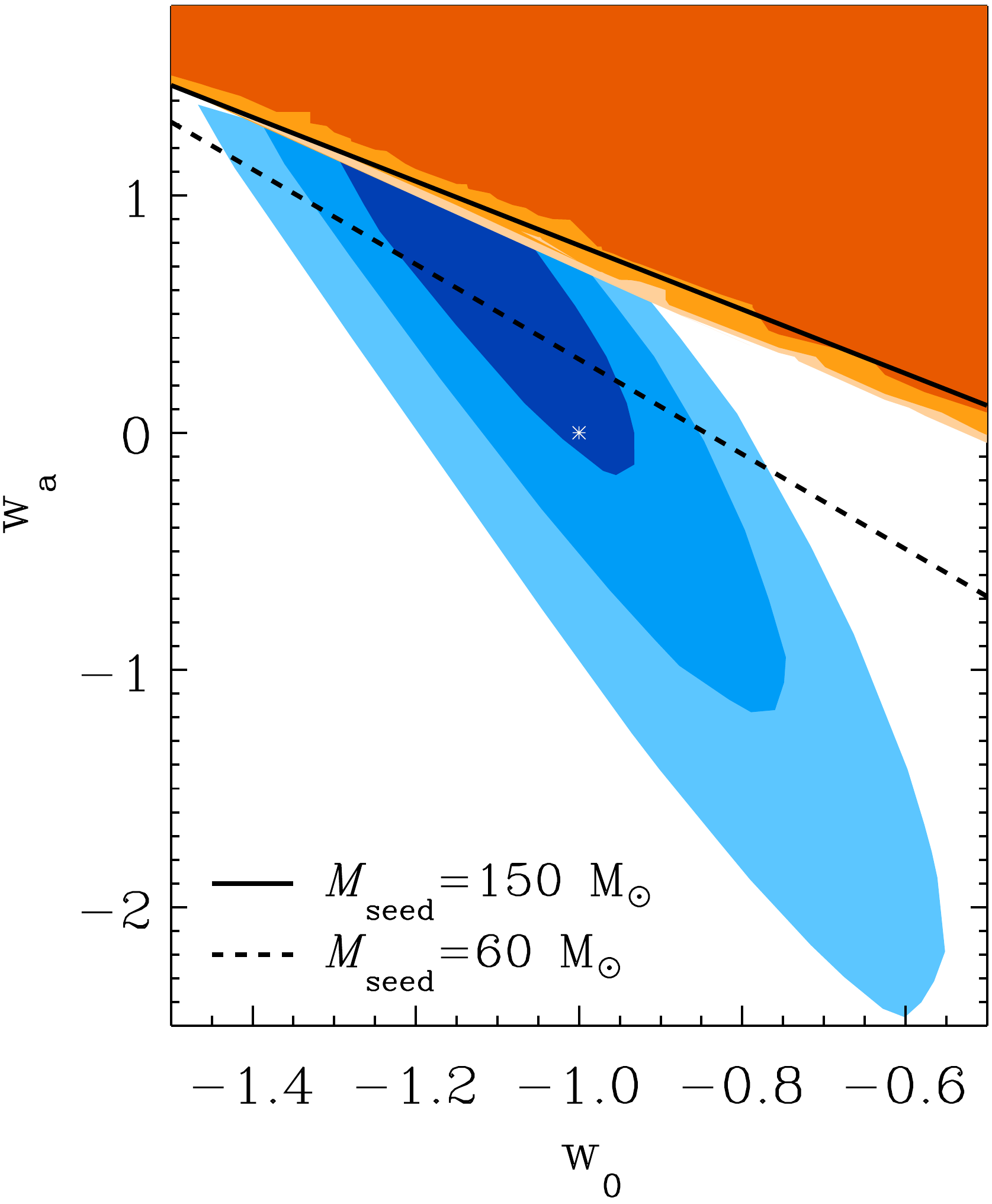}
\includegraphics[width=5 cm]{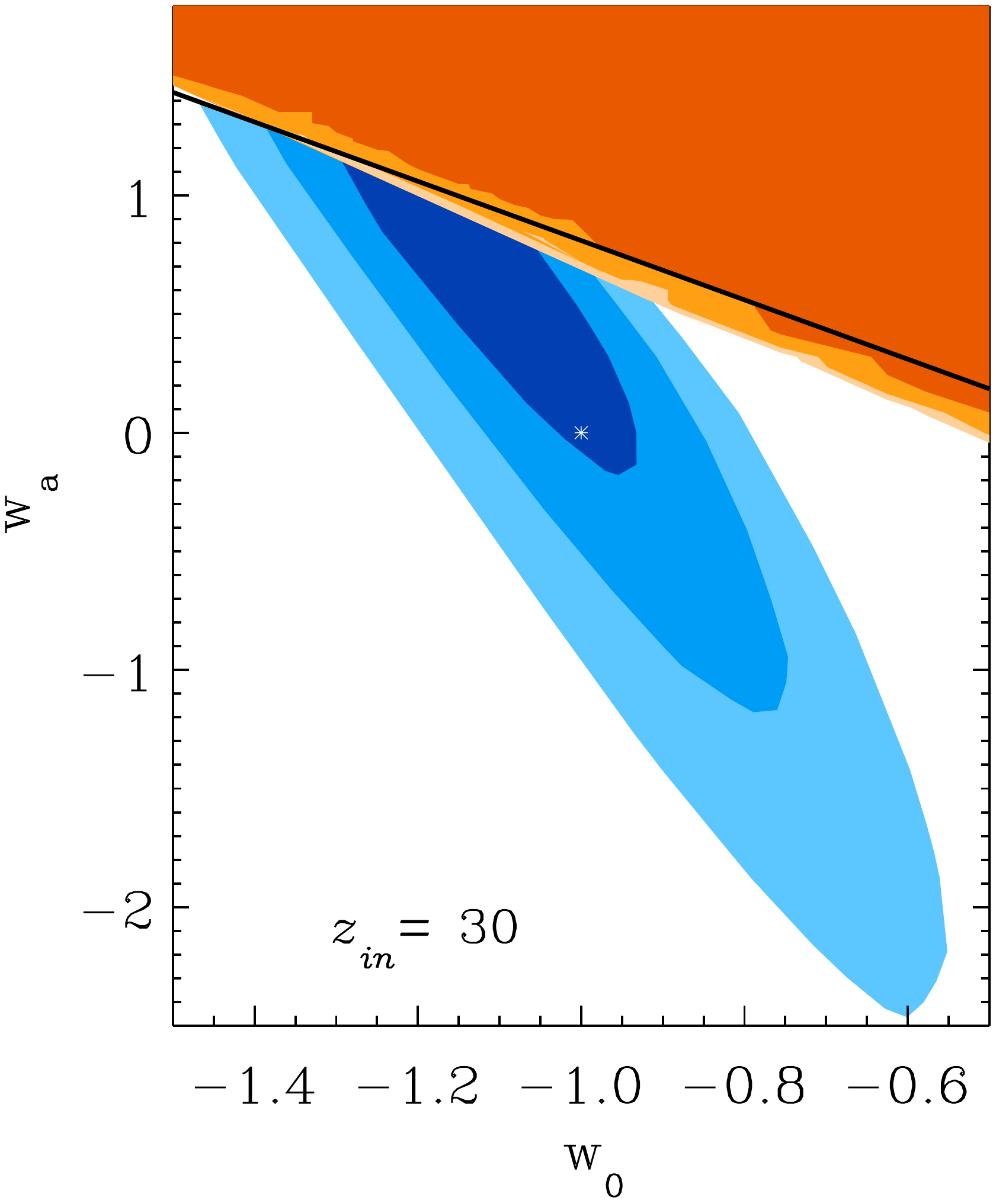}
\includegraphics[width=5 cm]{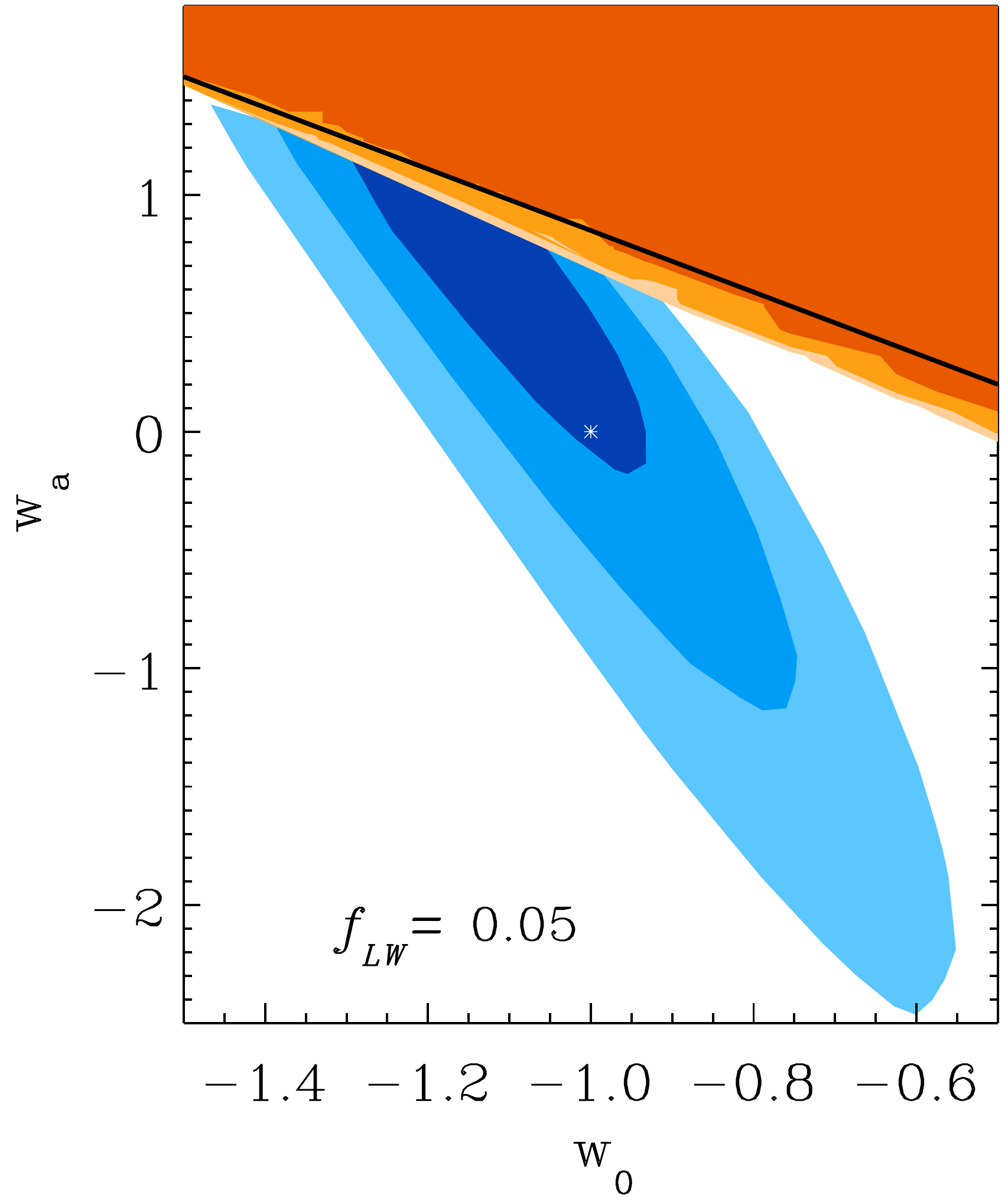}
\includegraphics[width=5 cm]{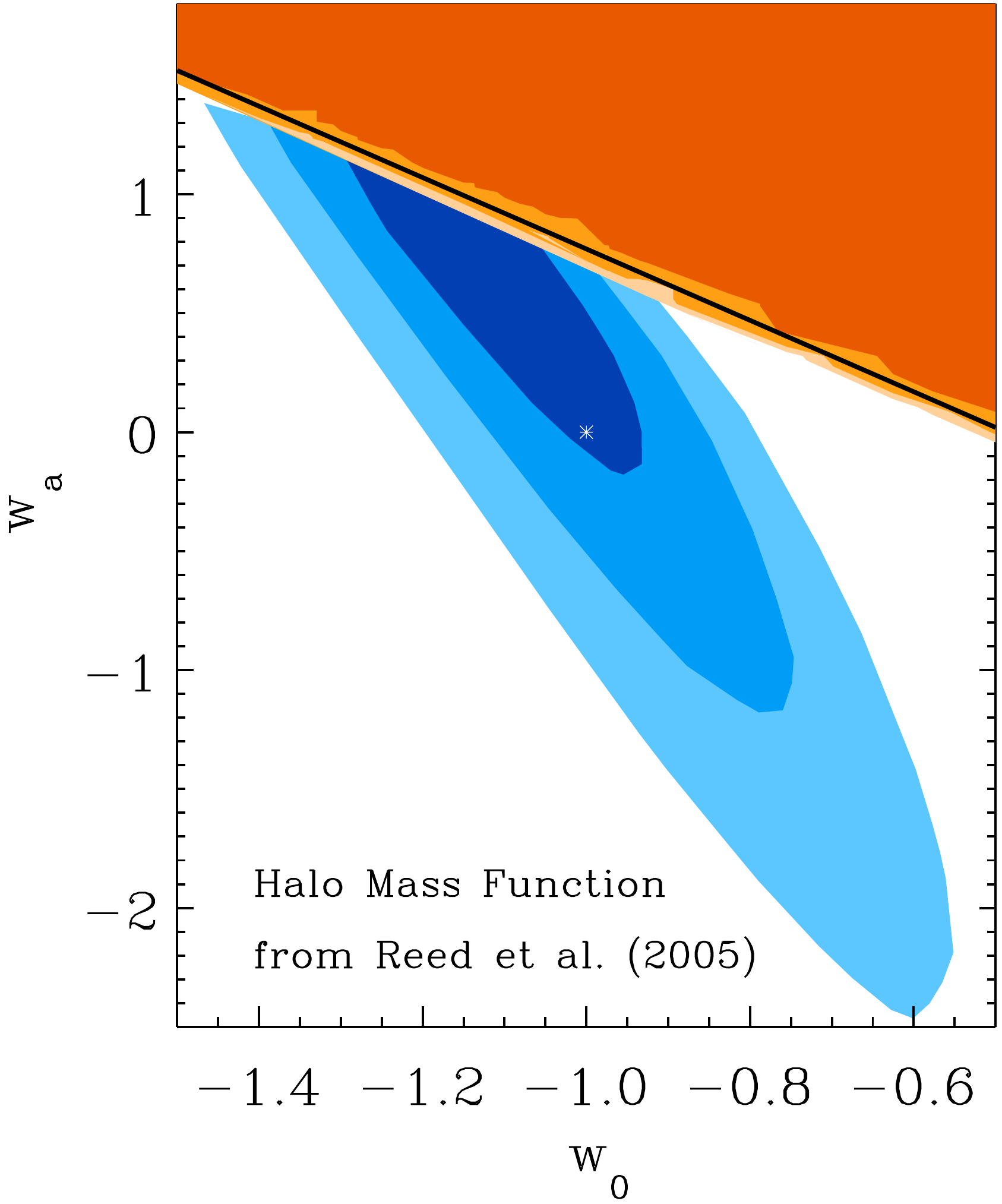}
\includegraphics[width=5 cm]{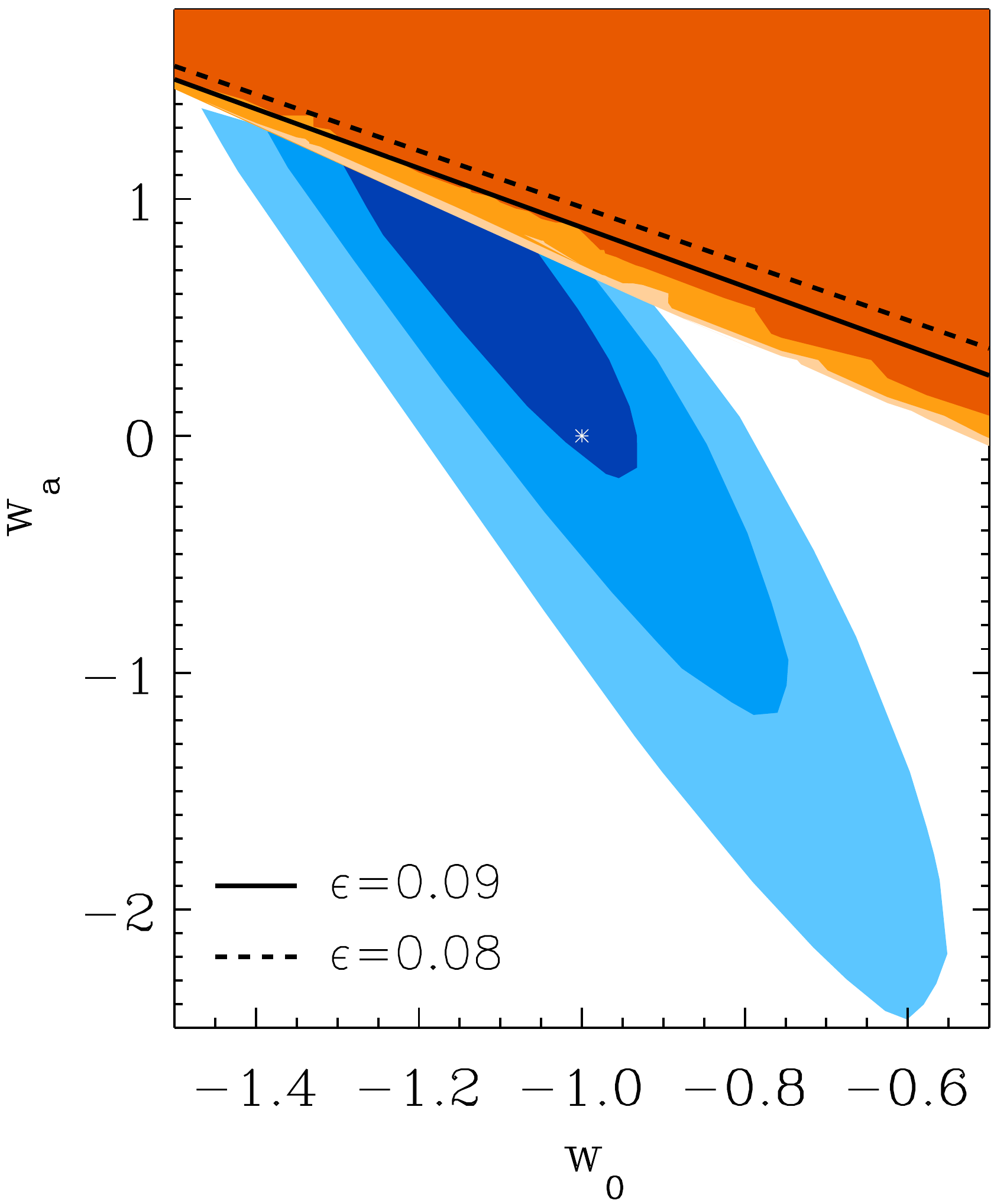}

\caption{Filled contours and shaded ellipses as in Fig 4. The lines represent the 1$\sigma$ exclusion regions in the $w_0$-$w_a$ plane derived assuming: M$_{seed}$ = 60 M$_{\odot}$ (Top Left panel, dashed line);  M$_{seed}$=150 M$_{\odot}$ (Top Left panel, solid line);  $z_{in}$=30 (Top Central panel);  $f_{LW}$=0.05 (Top Right panel); the halo mass function obtained by Reed et al. (2007, Bottom Left Panel); a BH accretion efficiency $\epsilon$=0.09 (solid line) and $\epsilon$=0.08 (dashed line, Bottom Right Panel).
}\label{inc_seed}
\end{center} 
\end{figure}

  We show in Fig. \ref{inc_seed} (top left panel) how the 1$\sigma$ exclusion region in the $w_0$-$w_a$ plane changes if  M$_{seed}$ = 60 M$_{\odot}$ (dashed line) and M$_{seed}$=150 M$_{\odot}$ (solid line) are assumed instead of M$_{seed}$=100 M$_{\odot}$ as in our fiducial model. It is interesting to note that assuming a  low massive BH seed increases significantly the number of DDE models excluded by our method, while increasing   M$_{seed}$ up to the maximal value allowed by theoretical models (150 M$_{\odot}$) does not reduce the exclusion region significantly. This is due to the fact that the decrease of  cosmic time and of  growth factor with increasing  $w_a$ is larger for high values of $w_a$ (see Fig. 1). In turn, this implies that the cosmological constraints derived in the previous section are not very sensitive to an increase of the BH seed mass  because in this region of the $w_0$-$w_a$ plane the increase of M$_{seed}$  is balanced by a small increment of $w_a$. It is also to be noted that the exclusion regions mainly differ at large values of $w_0$ because the decrease of  cosmic time and of growth factor with increasing $w_a$ are less prominent than at smaller values.

As for the uncertainties related to the formation redshift of seed BHs,  
theoretical models and cosmological simulations predict that  primordial star formation can take place in haloes at  30$\leq z_{in}\leq$20 (see Bromm et al. 2009 and reference therein). In our fiducial model we set $z_{in}$=20. Assuming that BH seeds are formed at higher redshift  would not change the cosmic density of seed BHs because it corresponds to populate  lower mass haloes with VMS  for fixed LW photon flux (for $z_{in}$=30, M$_{inf}$ decreases by a factor of about 20).
In contrast, increasing  the initial redshift  affects the BH growth due to gas accretion.
Infact, gas accretion onto the BH is determined by the ability of the halo to cool the gas efficiently. For atomic cooling this holds  for $M>10^{7.9}[(1+z)/20]^{-3/2} M_{\odot}$, corresponding to  $M>10^{8} M_{\odot}$ at $z=$20. Increasing the initial redshift up to $z_{in}$=30 requires accounting for the growth due to accretion of BH hosted in  haloes with $M>5\times 10^{7} M_{\odot}$; this slightly enhances the mass function of  BHs at $z$=6 as  illustrated in Fig.  \ref{inc_seed}  (central panel) by the small shift upwards of the 1$\sigma$ exclusion region (solid line).  This analysis demonstrates that our results are insensitive to the assumed value of the initial redshift. This validates our approach assuming that the initial redshift of seed BHs formation  is the same for all the DE cosmological models. 

The last parameter characterizing the BH seed population is the  minimum mass of the host DM halo. This is derived by a condition on the critical LW background flux ($f_{LW}$) above which  the H$_2$ cooling is prevented. Haiman, Abel \& Rees (2000) calculated  $f_{LW}$ as a function of the halo virial temperature and collapse redshift. They found that the higher the collapse redshift or temperature, the higher the flux needs to be to prevent star formation.  For  haloes with $10^{2.4}$ K $ \lesssim T_{vir} \lesssim 10^{3.8}$K, and for 10$<z<$50, they found values of the critical  background flux in the range $f_{LW} $=(10$^{-4}$-10$^{-2}$) photons per baryon, 
while for $T_{vir} \gtrsim 10^{4}$K the value of the flux is irrelevant because at these high temperatures cooling from neutral H always dominates over H$_2$. We conservatively adopted the upper limit of the above uncertainty range for $f_{LW}$ 
since lower values would yield even tighter constraints in the $w_0$-$w_a$ plane. However, even such a limit may be prone to some 
uncertainty: in fact, the density profile of the gas cloud is crucial in determining $f_{LW}$ , because the formation of H$_2$ molecules can be enhanced by the accelerated chemistry inside the central, dense regions. In addition, a central condensation tends to make the clouds more self-shielding against the external UVB background, and helps to preserve the internal H$_2$ molecules. The values of the critical flux mentioned above were obtained by Haiman et al. (2000) by assuming  a truncated isothermal sphere density profile  and by taking into account  self-shielding due to both H and H$_2$. In principle, different gas density profiles and  the self-shielding within haloes  could alter the estimate of the critical flux. It is difficult to quantify these effects,  doing this would require computing  radiative transfer across individual gas clouds exposed by an external UVB background, which is beyond the scope of this paper.  However, in order to explore the possible impact of such uncertainties on the $w_0$-$w_a$  constraints we repeated our calculations (for all the examined DE models) after increasing $f_{LW}$ by a factor of 5 above the already conservative fiducial value. 
The increase of the threshold of the LW flux up to $f_{LW}$=0.05 implies a lower minimum mass of the DM halo hosting primordial stars by about a factor of 2. This results in a higher number density of seed BHs present at the initial redshift,  and therefore a mild reduction of the DE models excluded by our  criterion ( see the solid line in the top-right panel of Fig. \ref{inc_seed}).

The final source of uncertainty related to the DM haloes hosting the seed BHs at the initial time is the halo mass distribution. In our fiducial model we assumed a  Press \& Schecter form (Sect. 2.1).  However, advanced N-body simulations indicate that PS may underestimate the real distribution in the high mass tail (e.g. Reed et al. 2007).  Since the most massive BHs are hosted in the most massive DM haloes it is important to show how  cosmological constraints change for a flatter DM halo mass function at the high-mass end . The bottom-left panel of Fig. \ref{inc_seed}  shows the 1$\sigma$ exclusion region in the $w_0$-$w_a$ plane derived  assuming the best-fitting formula for the halo mass function obtained by Reed et al. (2007) . Also in this case the exclusion region is reduced only by a small extent.

%Figure \ref{inc_seed} (right panel) shows the how the 1$\sigma$ exclusion region changes assuming $z_{in}$=30 (solid line). 
%for halo mass  $z>20/(M/10^{7.9})^{2/3}-1$ or equivalently
%however note that the mass of the halo where gas. can cool efficiently decreases as $z_{in}$ increases.
%The value of the formation redshift of seed BHs ($z_{in}$) determines the time interval in which black holes can grow through gas accretion. If $z_{in}$ increases, more time is left available for BH accretion in massive haloes where atomic cooling can take place,  corresponding to halo mass $M>10^{7.9}[(1+z)/20]^{-3/2} M_{\odot}$ 

%The increase of $z_{in}$ implies the formation of more massive BH at $z$=6  with implication on the cosmological model constraints. Figure \ref{inc_seed} (left panel) shows the 1$\sigma$ exclusion region in the $w_0$-$w_a$ plane derived assuming $z_{in}$=25 (solid line),  for comparison we also show the case $z_{in}$=19.\\

% while a decrease of these parameters enhances the number of discarded DDE cosmological models appreciably. This is due to the fact that for a given value of $w_0$ the decrease of the cosmic time and of the growth factor are larger for higher values of $w_a$ (see eq. (7) mettere in figura 1 DDE models che fanno vedere bene questo comportamento). It is also to be noted that the substantial variations are  for large values of $w_0$, which are outside the region  allowed by the other cosmological probes, because the decrease of the cosmic time and growth factor with the increase of $w_a$ are less pronounced at large values of  $w_0$. \\

We now  discuss our assumptions concerning BH accretion, i.e., radiative efficiency $\epsilon\leq 0.1$, and 
accretion not exceeding the Eddington value. In our fiducial model we conservatively fixed $\epsilon$=0.1, at the low end of the range where theoretical expectations overlap with observational constraints (see Sect. 2.2.2).  Although the constraints are derived from observations at redshifts $z<6$ and  the conditions  at $z>$6 could be different, all theoretical models and simulations concur in predicting that spin-up of BHs
should occur in gas-rich mergers between galaxies of comparable mass (Volonteri 2010; Dotti et al. 2010; Shapiro et al. 2005; see also Fanidakis et al. 2009). These are indeed the conditions expected to hold at higher redshifts in hierarchical models of galaxy formation (see, e.g., the review by Baugh et al. 2006 and references therein), which constitute the basic assumed framework of our paper. 

However, to investigate the change in our constraints on $w_0$-$w_a$ in response to the assumed limit on the $\epsilon$,  we repeated our computation  assuming
 $\epsilon<0.1$  for all DDE models. The solid and dashed lines in fig. \ref{inc_seed} show the 1$\sigma$ exclusion regions in the $w_0$-$w_a$ plane derived assuming $\epsilon=0.09$ and $\epsilon=0.08$ respectively.  Although a lower value of $\epsilon$ implies faster evolution of the BH mass function (and hence a smaller exclusion region for the DE cosmological models)  the basic conclusions drawn in Sect. 3.1 remain valid for $\epsilon>0.08$. For $\epsilon<0.08$ our approach does not allow to improve the limits on $w_0$-$w_a$ already provided by Supernovae, CMB and BAO;  note however that the observational limit $\epsilon\geq 0.08/\xi_0(1-\xi_m)$ discussed in Sect. 2.2.2 
would (marginally) allow such values $\epsilon< 0.08$ only when no merging is assumed (corresponding to $\xi_m=0$), a condition which is  not consistent with the hierarchical scenario of galaxy formation assumed in this work. Conversely, our constraints on $w_0$-$w_a$ shown in figs. 4 and 5 hold in all hierarchical scenarios where merging contributes to the growth of galaxies and BHs. 
 
As for the Eddington limited accretion assumption, we note that,  even if there is evidence for BH accretion above Eddington  in some sources, the 
effective accretion rate entering eq. (3) is the mean value (i.e., resulting from averaging the accretion rate over the statistical distribution of
rates). All existing observations are consistent with  an average value  below the Eddington limit both in the local (see, e.g., Netzer et al. 2009; 
Kauﬀmann \& Heckman 2009) and in the high redshift Universe (Trakhtenbrot et al. 2011; Shemmer et al. 2004; Netzer \& Trakhtenbrot 2007, Willott et al. 2010b). 
 We have also checked that our constraints are not weakened by uncertainties in the values of $H_0$ and $\Omega_M$ within the ranges allowed by observations.
Indeed, for all values of  $H_0$ consistent with SNe Ia at $z<$0.1 (Riess et al. 2009) , 
we obtain equal or tighter constraints on DDE models for all values of $\Omega_M$ allowed by the WMAP data for $w >-0.5$; the latter range includes  all the combinations 
$w_0-w_a$ for which our approach can provide significant limits (at the high redshifts $z\gg 1$ corresponding to the WMAP measurements  the Chevallier-Polarski-Linder parametrization yields $w\sim const \sim w_0+w_a$). 

The results shown in Fig. 5 indicate that  the cosmological constraints derived in this work are robust. Indeed, we have shown that  accounting for plausible uncertainties in the parameters describing the BH and DM halo populations 
does not imply a significant variation of the DE model exclusion region.  Furthermore, the exclusion regions  are mainly altered for large values of $w_0$, which are already excluded by  other cosmological probes.

%We  also investigate how these results change when the cosmological parameters  $\Omega_M$  ($\Omega_{\Lambda}$=1-$\Omega_M$) and $H_0$ vary within their observationally allowed range.  The WMAP data probes the high-redshift universe (z$\sim$1000) where the Chevallier-Polarski-Linder parametrization yields $w\sim cost \sim w_0+w_a$. The allowed values of $\Omega_M$ and $H_0$ derived only from the WMAP data (Larson et al. 2011)  assuming a costant value of $w\neq$-1 are $\Omega_M$=0.3-0.4 and $H_0$=50-60 Km/s/Mpc for w=-0.3, -0.2 which are the high-redshift value of the DDE models excluded by our criterion. 
%Assuming a larger value of $\Omega_M$ with respect $\Omega_M$=0.27 as in our fiducial model results in a decrease of the cosmic time a $z=$6 (eq. 8) . It also implies larger  growth factors (eq. 9 and 10), however the decrease of the  time available for the accretion of the BHs overwhelms this effects implying less massive BHs at $z=$6.  On the other hand, assuming a lower value of the Hubble costant  implies an increase of the cosmic time a $z$=6. We checked that allowing for a larger value of $\Omega_M$ ($\Omega_M$=0.3) in combination with a lower value of $H_0$ ($H_0$=50-60 Km/s/Mpc) leaves pratically unchanged the cosmological constraints we derived.

Finally, we checked our results on the BH mass function from eq. (3) against a full 
semi-analytic model (SAM) of galaxy formation. Such SAMs 
(see, e.g., Kauffmann \& Haenhelt 2000; Monaco \& Fontanot 2005; Menci et al. 2006; Croton et al. 2006; Bower et al. 2006; Marulli et al. 2008) are {\it ab-initio} numerical models of galaxy formation 
which relate the physical processes involving the 
baryons (gas physics, star formation, energy feedback from Supernovae onto the gas) to the collapse and the 
merging  histories of the DM haloes; the haloes are provided by  Monte Carlo simulations starting from 
the collapse of perturbations from the primordial density field. Galaxies form from dense gas cooling in the 
host DM haloes; when the latter merge to form larger haloes, the galaxies
may survive as satellites, merge to form larger galaxies,  or coalesce into a central dominant galaxy; these processes take place over time scales that grow longer over
cosmic time, so the number of satellite galaxies increases as the DM host haloes
grow from groups to clusters. Recent SAMs include specific physical models 
for BH accretion and hence for AGN activity. Here we compare with the  SAM by Menci et al. (2005, 2006, 2008) where galaxy interactions are assumed to trigger  the fueling of  black holes. Indeed, gas-rich mergers between galaxies of comparable mass have long been advocated to drive quasar activity by funnelling large amounts of galactic gas toward the galactic centre (see, e.g., Barnes \& Hernquist  1996; Cattaneo et al. 1999; Cavaliere, Vittorini 2000; Kauffmann, Haenhelt 2000; Wyithe, Loeb 2003; Treister et al. 2010). Such a picture is supported by hydrodynamical N-body simulations (Springel et al. 2005, Cox et al.  2008), 
which have shown that tidal torques during galaxy mergers can drive the rapid inflows of gas that are needed to fuel both the intense starbursts and rapid BH accretion associated with ULIRGS and QSOs (Hernquist 1989; Barnes 1992; Mihos \& Hernquist 1994; Barnes \& Hernquist 1996; Mihos \& Hernquist 1996; Di Matteo, Springel \& Hernquist 2005; Springel, Di Matteo \& Hernquist 2005a,b, 
Cox et al. 2008). Quasar activity triggered by mergers can account for the evolution of their luminosity function and for a wide range of QSOr properties at different wavelengths (Hopkins et al. 2005, 2006). Observations also support a link between merging and quasar activity (Sanders et al. 1988; Canalizo, Stockton 2001; Guyon et al. 2006; Dasyra et al. 2007; Bennert et al. 2008). 
%The merger-induced AGN activity is  commonly included in up-to-date semi-analytic models of galaxy formation (see Somerville et al. 2004b; Volonteri et al. 2003, 2006; Monaco \& Fontanot 2005; Bower et al. 2006;  et al. 2006; Menci et al. 2003, 2006, Croton et al. 2006, Hopkins  et al. 2006). 
In the Menci et al. (2005, 2006, 2008)  models the interaction rate for galaxies 
in a common DM halo is calculated as: $\tau_r^{-1}$=$n\,\Sigma\,V_{rel}$, where $n$ is the number density of galaxies in the same halo,
$V_{rel}$ is their relative velocity, and $\Sigma$ is the cross section for
such encounters. The amount of cold gas accreted during an accretion episode
is: $\Delta m_{acc}=f_{acc} m_{c}$, where $m_{c}$ is the amount of cold gas in
the galaxy disk (with radius $r_d$ and rotational velocity $v_d$);
$f_{acc}$ is the fraction of the cold gas destabilized in an interaction event,
which is computed in terms of the variation of the specific angular momentum
of the gas in the disk (Menci et al. 2003). The duration of an accretion
episode is assumed to be the crossing time for the destabilized cold gas
component: $\tau$=$r_d$/$v_d$. At high redshifts, large values of $m_{c}$ and $f_{acc}$ are
obtained (corresponding to effective BH accretion close to the Eddington limit).
The first stems from the rapid gas cooling owing to the low  virial
temperatures and to the large densities of the haloes at early cosmic times.
The second is due to the comparable galaxy masses involved in high-$z$ interactions. 

The BH mass function computed from the full SAM is compared with our solutions of eq. (3) in Fig. 5  for the 
$w=1$ case (the only cosmology for which SAM results are available); note that our approach based on eq. (3) not only 
assumes continuous BH accretion at the Eddington limit, but also maximizes the BH growth  by assuming that the BH merging immediately follows
the halo merging, and that the BHs are always accreting at the Eddington limit. 
Thus, the fact that the our solutions to eq. (3) lie always {\it above} the SAM predictions constitutes 
an important consistency check for our approach, reinforcing the robustness of our constraints on the DDE models. 

The different shape of the BH mass functions at the 
low- and at the high-mass end has interesting physical explanations, 
which enlighten how our results in Sect. 3 actually maximized the evolution of the 
BH growth. For large masses $M_{BH}\gtrsim$10$^{9}$ $M_{\odot}$ the SAM predicts lower BH number
densities because: i) in the SAM, BH merging does not immediately follow the coalescence of DM haloes, since the dynamical friction times become too long in massive haloes; ii) the gas cooling is less efficient due to large halo virial temperatures, 
and this reduces the amount of cold gas available for the BH accretion.  At the 
low-mass end $M_{BH}<$10$^{8}$ $M_{\odot}$ the lower BH number densities predicted by the
SAM are due to the feedback from Supernovae exploding in the host galaxy. Feedback expels/reheats part of the 
cold gas in the shallow potential wells, thus decreasing the reservoir of cold galactic gas available for BH accretion. Note however that 
for BH masses $10^8\,M_{\odot}\leq M_{BH}\leq 10^9\,M_{\odot}$ the shape of the BH mass function is insensitive to these processes.

% and the its slope actually reflects the effect of the AGN duty cycle, as discussed above.\\
%Observationally, the shape of the mass function is not strongly constrained below $M_{BH}$ $\simeq$10$^{9}$ $M_{\odot}$. Indeed,  present optical surveys such as the SDSS (Fan et al. 2006, Jiang et al. 2009) and the CFHQS (Willott et al. 2010a) have been able to discover $\sim$ 50 quasar at $z$=5.7-6.4 (squares and triangles in Fig. \ref{BHMF_SAM}), however  this surveys are biased against low-luminosity and obscured sources. 
%implying that the derived mass function is not strongly constrained below $M_{BH}$ $\simeq$10$^{9}$ $M_{\odot}$ .  
%Lower luminosity and/or obscured AGN at high $z$ can be detected directly in current X-ray  surveys (Fiore 2011). In fact, hard X-ray selection provides a more complete and direct sampling of the AGN population, being less biased against obscured sources. 
%In addition, while in the optical  the K-correction works against the detection of obscured sources (the dust extinction being larger at higher frequencies),  at $z$=6 the X-ray band observed by $Chandra$ and $XMM-Newton$ (0.5-7 keV) corresponds to 3.5-42 keV rest frame, where the effect of obscuration is less important; in fact, the X-ray spectrum is not affected by the sharp cut-off  (which depends on the column density of the absorber) due to the photoelectric absorption of the primary emission. 
  
\begin{figure}
\begin{center}
\includegraphics[width=8 cm]{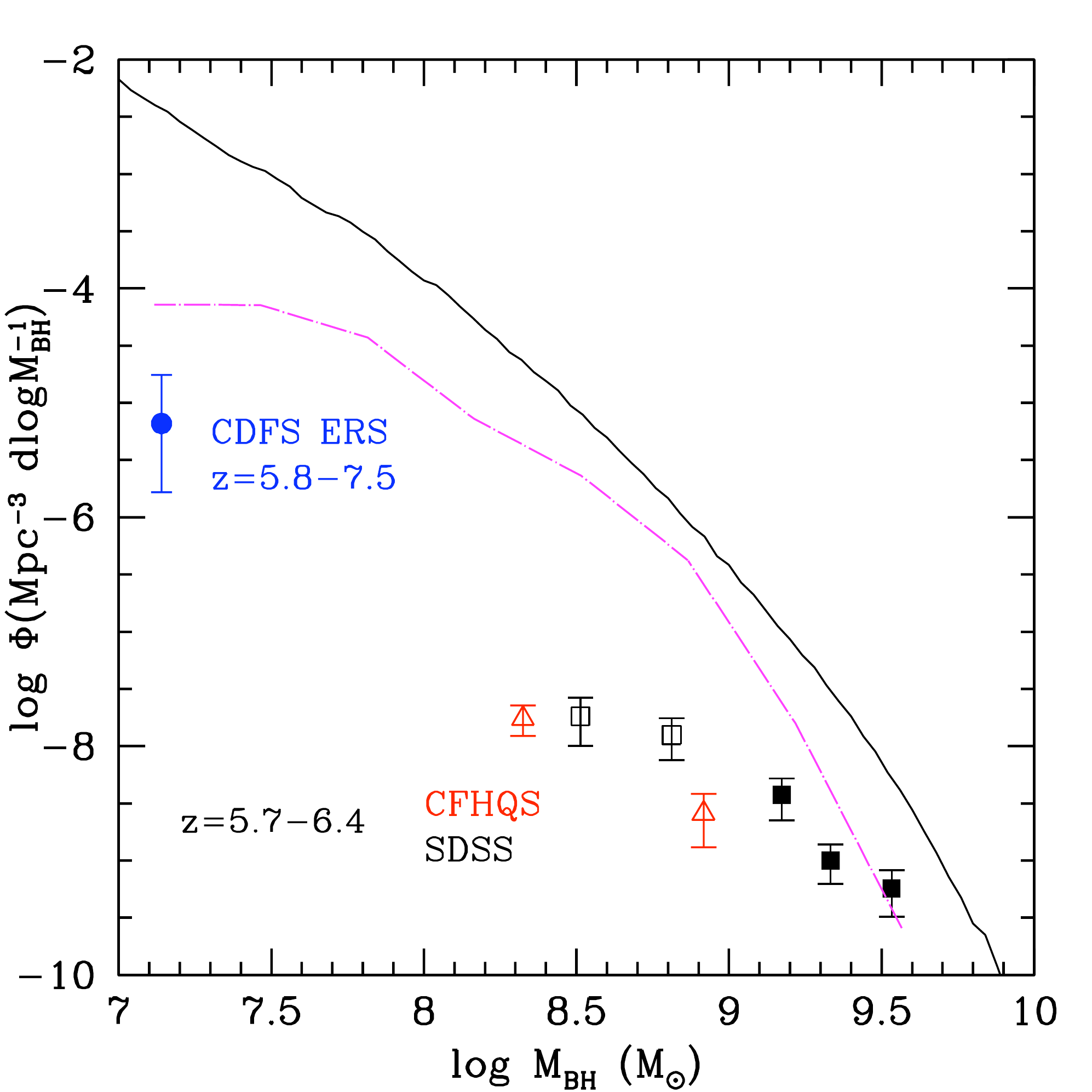}
\caption{The black hole mass function at $z$=6, in the $\Lambda$CDM cosmology, derived in this work (solid line) is compared with the BH mass function predicted by the SAM developed by Menci et al. (2005, 2006, 2008) (dot-dashed line). Squares and triangles represent the the BH number densities which we derive from the SDSS and CFHQS surveys respectively; the filled circle is the point derived by Fiore et al. (2011). Open symbols indicate the region where the mass function is poorly constrained.}
\label{BHMF_SAM}
\end{center} 
\end{figure}

%\end{mathletters}
\section{Summary \& Conclusions}

We have computed the number density of massive BH at the centre of
galaxies at $z=6$ in different DDE cosmologies; from the comparison with 
existing observational lower limits we derived constraints on the DE
equation of state $w$. Our approach only assumes the
canonical scenario for structure formation from the collapse of
overdense regions of the DM dominated primordial density field on
progressively larger scales; the BH accretion and merging rate have
been {\it maximized} in the computation so as to obtain robust constraints
on the normalization and on the time evolution of $w$ that we
parametrized as $w=w_0+w_a(1-a)=w_0+w_a\,z/(1+z)$, following Linder
(2003). This expression provides an effective fitting formula for the
time evolution of the equation of state of a number of physically
motivated DE fields.

Our results shown in Fig. 4 put novel constraints on $w_0$ and $w_a$
which are remarkably complementary to previous results from combined
BAO, SNae and CMB experiments. In fact, the latter mainly involve
either integrals of the Hubble expansion rate $H(a)$ over a large
cosmic time (from $z\approx 1000$ to the present) or the value of
$H(a)$ at relatively low redshifts $z\lesssim 1$,
thus providing constraints on  $w_0$ and $w_a$ which are strongly correlated.
On the contrary, our results concerning the number density of massive BH
are effective in constraining all DDE models which do not provide
cosmic ages and fast growth factor (Fig. 1 and 3) large enough to
allow for the building up of the observed abundance of massive BHs at $z=6$;
this corresponds to setting constraints mainly on the 
time evolution $w_a$ {\it at high redshifts}, almost regardless of $w_0$. In
particular, models with positive values of $w_a$ (corresponding to a
positive redshift evolution of $w$) can be effectively constrained by our 
approach, since the DE density in $H(a)$ (eq. 7) entering the ages
(eq. 8) increase with $w_a$ like $a^{-3(1+w_0+w_a)}e^{3w_a(a-1)}$;
larger DE densities at earlier epochs also imply a delay in the growth
of DM perturbations resulting in a slower growth factor.
On the basis of a conservative approach to
compute the {\it maximal} BH abundance at $z=6$, we strongly disfavour
models represented by the upper
area in Fig. 4; an useful fitting formula for such an exclusion region 
is  $w_a\gtrsim -3/2\,w_0-3/4$. 
In particular, models with $-1.2\leq w_0\leq
-1$ and $w_a\gtrsim 0.8$ - completely consistent with previous
constraints (and indeed mildly favoured according to some recent
analysis, see Feng, Wang, Zhang 2005; Upadhye, Ishak, Steinhardt 2005;
and the analysis of the "Gold" dataset in Perivolaropoulos 2006) - are
excluded by our results. Such range of parameters corresponds to
"Quintom" DDE models, with $w$ crossing $-1$ starting from larger
values, a transition that cannot be fulfilled by pure Quintessence or
Phantom fields.

For Quintessence models with $w\gtrsim -1$, our results exclude DDE
with an equation of state rapidly evolving with $z$, so that
$-dw/da\gtrsim -3/2\,w_0-3/4$. Such an evolution of the equation of
state can be related to a class of dynamics of the DE field $\phi$. In
fact, the latter is given by the Klein- Gordon equation $\ddot
\phi=-V_{,\phi}-3\,H\,\dot\phi$, where the first term on the r.h.s. is
positive and represents the contribution due to the steepness of the
potential $V_{,\phi}\equiv dV/d\phi$, while the second is negative and
corresponds to the friction due to the Hubble expansion. The coasting
evolution $\ddot \phi=0$ corresponds to an unstable situation of
perfect balance between friction and the potential terms, dividing the
dynamics into a class of "thawing" solutions with $\ddot \phi>0$ and a
class of "freezing" solutions with $\ddot \phi<0$. The first
corresponds to dynamics initially dominated by the friction term, with
the field evolving away from a "cosmological constant-like" state,
while the second to DE fields decelerating as they evolve down their
potential toward the minimum, since the forcing due to the steep slope
of the potential dominates over the Hubble drag at early times; as the
minimum is approached, the flatter shape of the potential leaves the
Hubble drag  dominate the dynamics, effectively freezing the field.
The coasting line $\ddot \phi=0$ separating the two behaviours
corresponds to the condition $dw_a/da=3(1-w^2)/a$ Linder (2006); thus
the number density of high-redshift BHs has an impact in excluding a
wide class of models with a "freezing" behaviour corresponding to
initially steep potentials. Indeed, SUGRA inspired models (Brax \&
Martin 1999), which are well fitted by $w_0\approx -0.82$ and
$w_a\approx 0.58$ (Linder 2003) are also highly unfavoured due to the
combination of our constraints with previous observational
constraints, as results from of Fig. 4.

The cosmological constraints we derived are based on the standard gravity theory, the same approach will be applied to modified gravity models (e.g. the Dvali, Gabadadze, Porrati 2000 model ) in the next future.

Note that the exclusion regions we derived in the $w_0-w_a$ plane are
extremely conservative, since in our computation we maximized the
effectiveness of all processes contributing to the growth of BHs. In
particular the above constraints are derived assuming: i) continuous
BH accretion (no time gap between successive accretion episodes); ii)
accretion always at the maximal Eddington limit; iii) BH merging
immediately following the merging of their host DM haloes; iv) all
baryons contained in the galaxy are available for BH accretion and no
energy feedback balancing the gas cooling in DM haloes. Such
conditions are expected to be exceedingly restrictive, as we have
shown in Sect. 4 using a semi-analytic model of galaxy formation
models based on a physical description of BH accretion.

The above conservative approach makes difficult to extend the present 
method to lower redshifts, since our maximal accretion assumptions extrapolated 
over a larger span of cosmic time would 
yield, for all plausible $w_0$-$w_a$ combinations,  upper bounds 
largely exceeding the observed abundances, and thus ineffective for constraining  
DDE models. Conversely, our approach is even more effective if the upper bounds  
derived from eq. 3 could be compared with the measured abundances of QSOs at 
higher redshifts $z>6$.  In this context,  
the discovery of three new quasars at $z\simeq$7 in the United Kingdom Infrared Deep Sky Survey  
(UKIDSS, Mortlock et al. 2011;  Venemans et al. in prep.) 
indicate that the present approach will be able to 
provided even tighter constraints on DDE models in the next future.

\section*{Acknowledgments} We acknowledge grants from ASI-INAF I/016/07/0 and ASI-INAF 1/009/10/0. We thank the referee for helpful comments.

%\bibliographystyle{mn2e}
%\bibliography{biblio_dde}

\begin{thebibliography}{}

\bibitem{}Abel T., Bryan, G.L., Norman M.L., 2000, ApJ, 540, 39
\bibitem{}Abel T., Bryan G. L., Norman M. L., 2002, Science, 295, 93
\bibitem{}Alcaniz J.S., 2004, Phys. Rev. D, 68, 083521
\bibitem{}Amanullah, R. et al. 2010, ApJ, 716, 712
\bibitem{}Amara A., Refregier A., 2007, MNRAS, 381, 1018
\bibitem{}Amendola L., 2004, Phys. Rev. D, 70, 103522
\bibitem[\protect\citeauthoryear{{Ascasibar} \& {Gottl{\"o}ber}}{{Ascasibar} \&
 {Gottl{\"o}ber}}{2008}]{Ascasibar08}
{Ascasibar} Y.,  {Gottl{\"o}ber} S.,  2008, \mnras, 386, 2022
\bibitem{}Bahcall N. A. et al., 2003, ApJ, 585, 182
\bibitem{}Barnes J.E., 1992, ApJ, 393, 484
\bibitem{}Barnes J.E., Henrquist  L., 1996, ApJ, 471, 115
\bibitem{}Bardeen J.M., Petterson J.A., 1975, ApJ, 195, L65
\bibitem{}Barth A.J., Martini P., Nelson C.H., Ho L.C., 2003, ApJ, 594, L95
\bibitem{}Baugh C.M., 2006, Rep. Prog. Phys. 69, 3101
\bibitem{}Begelman M. C., Volonteri M., Rees M. J., 2006, MNRAS, 370, 289
\bibitem{}Bennert N., Canalizo G., Jungwiert B., Stockton A., Schweizer F., Peng C.
Y., Lacy M., 2008, ApJ, 677, 846
\bibitem{}Berti E., Volonteri M., 2008, ApJ, 684, 822
\bibitem{}Bianchi, E., Rovelli, C., Kolb, R. Nature 466, 321–322
\bibitem{}Bogdanovic T., Reynolds C.S., Miller M.C., 2007, ApJ, 661, L147
\bibitem[\protect\citeauthoryear{{Bond}, {Cole}, {Efstathiou} \&
 {Kaiser}}{{Bond} et~al.}{1991}]{Bond91}
{Bond} J.~R.,  {Cole} S.,  {Efstathiou} G.,    {Kaiser} N.,  1991, \apj, 379,
 440
% \bibitem{}Borgani S., 2006, preprint (astro-ph/0605575 )
\bibitem{}Borgani S., Guzzo L., 2001, Nature, 409, 39
\bibitem[\protect\citeauthoryear{{Bower}, {Benson}, {Malbon}, {Helly}, {Frenk},
 {Baugh}, {Cole} \& {Lacey}}{{Bower} et~al.}{2006}]{2006MNRAS.370..645B}
{Bower} R.~G.,  {Benson} A.~J.,  {Malbon} R.,  {Helly} J.~C.,  {Frenk} C.~S.,
 {Baugh} C.~M.,  {Cole} S.,    {Lacey} C.~G.,  2006, \mnras, 370, 645
\bibitem{}Bower, R. 1991, MNRAS, 248, 332
\bibitem{}Brax, P. \& Martin, J., 1999, Phys.Lett., B468, 40  
\bibitem{}Bromm V.,Coppi P.S., Larson B.B., 1999, ApJL, 527, L5
\bibitem{}Bromm V., Coppi P.S., Larson B.B., 2002, ApJ, 564, 23
\bibitem{}Bromm V., Loeb, A. 2003, ApJ, 596, 34
\bibitem{}Bromm V., Yoshida M, Hernquist L., McKee C. F., 2009, \nat, 459,  49
\bibitem{}Callegari S. et al., 2009, ApJ, 696, L89
\bibitem{}Caldwell R.R., Dave R.,  Steinhardt P.J., 1998, Phys. Rev. Lett. 80 1582
\bibitem{}Caldwell R.R., Kamionkowski M., Weinberg N.N., 2003, Phys. Rev. Lett., 91, 071301
\bibitem{}Caldwell R.R., Linder, E.V 2005, Phys. Rev. Lett., 95, 141301
\bibitem{}Canalizo G., Stockton A., 2001, ApJ, 555, 719
\bibitem{}Cappelluti N. et al. 2010, Proceedings of the WFXT meeting, to appear on
"Memorie della Societa' Astronomica Italiana", arXiv:1004.5219
\bibitem{}Carrol S.M., Press W.H., Turner E.L., 1992, ARA\&A, 30, 499
\bibitem{}Carrol S.M., Hoffman M., Trodden M., 2003, 68, 043509
\bibitem{}Cattaneo A., Haenhelt M.G., Rees M., 1999, MNRAS, 308, 77
\bibitem[\protect\citeauthoryear{{Cavaliere} \& {Vittorini}}{{Cavaliere} \&
 {Vittorini}}{2000}]{Cavaliere00}
{Cavaliere} A.,  {Vittorini} V.,  2000, \apj, 543, 599
\bibitem{} Chevallier  M., Polarski D., 2001, Int. J. Mod. Phys. D 10,213
\bibitem{}Cole S. et al., 2005 MNRAS 362 505–34
\bibitem{}Coles P., Lucchin F., 2002 Cosmology: The Origin and Evolution of Cosmic Structure (New York: Wiley)
\bibitem{}Colpi M., Dotti M., 2009, Advanced Science Letters, in press
\bibitem{}Copeland E.J., Sahni M., Tsujikawa S.,  2006, Int. J. Mod. Phys.,  D15, 1753
\bibitem{} Cox T. J., Jonsson P., Somerville R. S., Primack J. R., Dekel A., 2008, MNRAS, 384, 386
\bibitem{} Croton D. J. et al.,  2006, \mnras, 365, 11
\bibitem{}Dasyra K. M. et al., 2007, ApJ, 657, 102
\bibitem{}Davis T. M. et al., 2007, ApJ, 666, 716
%\bibitem{}de La Calle P\'erez I., 2010, A\&A, 524, 50
\bibitem{}De Lucia, G. et al. 2004, MNRAS, 348, 333
\bibitem{}Dijkstra M., Haiman Z., Mesinger A., Wyithe S., 2008, MNRAS, 391, 1961
\bibitem[\protect\citeauthoryear{{Di Matteo}, {Springel} \& {Hernquist}}{{Di
 Matteo} et~al.}{2005}]{2005Natur.433..604D}
{Di Matteo} T.,  {Springel} V.,    {Hernquist} L.,  2005, \nat, 433, 604
\bibitem{}De Bernardis P. et al., 2000, Nature 404 955–9
\bibitem[\protect\citeauthoryear{{Diemand}, {Kuhlen} \& {Madau}}{{Diemand}
 et~al.}{2007}]{2007ApJ...667..859D}
{Diemand} J.,  {Kuhlen} M.,    {Madau} P.,  2007, \apj, 667, 859
\bibitem{}Dotti M., Colpi M., Hardy F., 2006, MNRAS, 367, 103
\bibitem{}Dotti M., Colpi M., Hardy F., Mayer L., 2007, MNRAS, 379, 956
\bibitem{}Dotti, M., Volonteri, M., Perego, A., Colpi, M., Ruszkowski, M., Haardt, F. 2010, MNRAS, 402, 682
\bibitem{}Dunkley J. et al., 2009, ApJ, 701, 1804
\bibitem{}Eisenstein D.J. et al., 2005, ApJ, 633, 560
\bibitem{}Elvis M., Risaliti G., Zamorani G., 2002,ApJ, 565, L75
\bibitem{}Fan X. et al., 2000, AJ, 120, 1167
\bibitem{}Fan X. et al., 2001, AJ, 122, 2833
\bibitem{}Fan X. et al., 2003, AJ, 125, 1649
\bibitem{}Fan X. et al., 2004, AJ, 128, 515
\bibitem{}Fan X. et al., 2006, AJ, 131, 1203
\bibitem{}Fanidakis N. et al., 2010, MNRAS, in press
\bibitem{}Feng B., Wang. X., Zhang X., 2005, Phys. Rev. Lett., B607, 35
\bibitem[\protect\citeauthoryear{{Ferrarese} \& {Merritt}}{{Ferrarese} \&
 {Merritt}}{2000}]{2000ApJ...539L...9F}
{Ferrarese} L.,  {Merritt} D.,  2000, \apjl, 539, L9
\bibitem{}Fiore F. et al. 2011, A\&A, in press, arXiv:1109:2888
%\bibitem{}Fiore F. et al. 2011, A\&A submitted
\bibitem{}Frieman J.A., Turner S.,  Huterer D.,  2008, ARA\&A, 46, 385
\bibitem{}Fryer C., Woosley S.E., Heager A., 2001, ApJ, 550, 372
\bibitem{}Fuller T.M., Couchman H.P.M., 2000, ApJ, 544, 6
\bibitem{}Gallerani S. et al., 2010, A\&A, 523, 85
\bibitem{}Gao L., Yoshida N., Abel T., Frenk C.S., Jenkins A., Springel V., 2007, MNRAS, 378, 449
\bibitem{} Gebhardt K. et al., 2000, \apjl, 539, L13
\bibitem{}Gehrels N., 2010, Bulletin of the American Astronomical Society, Vol. 42, p.590
\bibitem{}Gottl\"ober, S., Klypin, A., Kravtsov, A.V. 2001, ApJ, 546, 223
\bibitem{}Guo Z. et al., 2005, Phys. Lett. B, 608, 177
\bibitem{}Guyon O., Sanders D. B., Stockton A., 2006, ApJS, 166, 89
\bibitem{}Guzzo L. et al., 2008, \nat, 451, 541
\bibitem{}Haiman Z., Rees M.J., Loeb A., 1997,  ApJ, 476, 458
\bibitem{}Haiman, Z., Abel, T., Rees, M.J. 2000, ApJ, 534, 11
\bibitem{}Hanany S. et al., 2000, ApJ, 545, L5
\bibitem{}Hernquist L., 1989, Nature, 340, 687
\bibitem{}Hinshaw G. et al., 2003 Astrophys. J. Suppl. Ser. 148 135
\bibitem[\protect\citeauthoryear{{Hopkins}, {Hernquist}, {Cox}, {Di Matteo},
 {Robertson} \& {Springel}}{{Hopkins} et~al.}{2005}]{2005ApJ...632...81H}
{Hopkins} P.~F.,  {Hernquist} L.,  {Cox} T.~J.,  {Di Matteo} T.,  {Robertson}
 B.,    {Springel} V.,  2005, \apj, 632, 81
\bibitem[\protect\citeauthoryear{{Hopkins}, {Hernquist}, {Cox}, {Di Matteo},
 {Robertson} \& {Springel}}{{Hopkins} et~al.}{2006}]{2006ApJS..163....1H}
{Hopkins} P.~F.,  {Hernquist} L.,  {Cox} T.~J.,  {Di Matteo} T.,  {Robertson}
 B.,    {Springel} V.,  2006, \apjs, 163, 1
\bibitem{}Hopkins P.F., Richards G.T., Hernquist L.,  2007, ApJ, 654, 731
\bibitem{}Huterer D.V.,  Cooray A., 2005, Phys. Rev. D, 71, 3506
\bibitem{}Ivezic Z. ,2010, Bulletin of the American Astronomical Society, Vol. 42, 217
\bibitem{}Jiang L. et al., 2006, AJ, 132, 2127
\bibitem{}Jiang L. et al., 2009, AJ, 138, 305
\bibitem[\protect\citeauthoryear{{Kauffmann} \& {Haehnelt}}{{Kauffmann} \&
 {Haehnelt}}{2000}]{2000MNRAS.311..576K}
{Kauffmann} G.,  {Haehnelt} M.,  2000, \mnras, 311, 576
\bibitem{}Kauﬀmann G., Heckman T.M. 2009, MNRAS, 397, 135 
\bibitem{}Kesden M., Sperhake U., Berti E., 2010, Phys. Rev., D81, 084054
\bibitem{}King A.R., Pringle  J.E., 2006, MNRAS, 379, L80
\bibitem{}King A.R., Pringle  J.E., Hofmann  J.A., 2008, MNRAS, 363, 49
\bibitem{}Komatsu E. et al., 2008, ApJ, 180, 330
\bibitem{}Kowalski M. et al., 2008, Apj, 686, 749
\bibitem{}Kudritzki R.P., Puls  J., 2000, ARA\&A, 38, 613
\bibitem{}Kurk J. D. et al., 2007, ApJ, 669, 32
\bibitem[\protect\citeauthoryear{{Lacey} \& {Cole}}{{Lacey} \&
 {Cole}}{1993}]{1993MNRAS.262..627L}
{Lacey} C.,  {Cole} S.,  1993, \mnras, 262, 627
%\bibitem{}Lapi  A., Shankar F., Mao J., Granato G. L., Silva L., De Zotti G., Danese L.,
2006, ApJ, 650, 42
\bibitem{} Larson D. et al., 2011, ApJS, 192, 16
\bibitem{}Laurejis, R. 2008, EUCLID CDF Study Report: CDF-73(A) 
\bibitem{}Li H., Xia  J.-Q., Fan Z., Zhangm X., 2008, Cosmol. Astropart. Phys., 10, 46
\bibitem{}Linder E.V.,  2003, Phys. Rev. Lett. 90 091301
\bibitem{}Linder, E.V., Jenkins, A. 2003, MNRAS, 346, 573
\bibitem{}Linder E.V.,  2005, Phys. Rev. D,  72, 043529
\bibitem{}Linder E.V.,  2006, Phys. Rev. D,  73, 063010
\bibitem{}Linder E.V., Cahn R. N., 2007, Astropart. Phys., 28, 481
\bibitem{}Liu F.K., 2004, MNRAS, 347, 1357
\bibitem{}Lodato G., Natarajan P., 2007, MNRAS, 377, L64
\bibitem{}Madau P., Rees M., 2001, ApJ, 551, L27
\bibitem{}Mainini, R., Maccio, A.V., Bonometto, S.A., Klypin,  A. 2003, ApJ, 599, 24
\bibitem{}Maiolino R., 2004, IAUSS, 222, 229
\bibitem[\protect\citeauthoryear{{Marconi}, {Risaliti}, {Gilli}, {Hunt},
 {Maiolino} \& {Salvati}}{{Marconi} et~al.}{2004}]{2004MNRAS.351..169M}
{Marconi} A.,  {Risaliti} G.,  {Gilli} R.,  {Hunt} L.~K.,  {Maiolino} R.,
 {Salvati} M.,  2004, \mnras, 351, 169
\bibitem{} Marulli, F., Bonolis, S., Branchini, E., Moscardini, L., Springel, V. 2008, MNRAS, 385, 1846
\bibitem{}Massey R. et al.,  2007, Nature 445, 286
\bibitem{}McKee C. F., Tan J., 2008, ApJ, 681, 771
\bibitem[\protect\citeauthoryear{{Menci}, {Cavaliere}, {Fontana}, {Giallongo},
 {Poli} \& {Vittorini}}{{Menci} et~al.}{2003}]{2003ApJ...587L..63M}
{Menci} N.,  {Cavaliere} A.,  {Fontana} A.,  {Giallongo} E.,  {Poli} F.,
 {Vittorini} V.,  2003, \apjl, 587, L63
\bibitem[\protect\citeauthoryear{{Menci}, {Fontana}, {Giallongo} \&
 {Salimbeni}}{{Menci} et~al.}{2005}]{2005ApJ...632...49M}
{Menci} N.,  {Fontana} A.,  {Giallongo} E.,    {Salimbeni} S.,  2005, \apj,
 632, 49
\bibitem[\protect\citeauthoryear{{Menci}, {Fontana}, {Giallongo}, {Grazian} \&
 {Salimbeni}}{{Menci} et~al.}{2006}]{2006ApJ...647..753M}
{Menci} N.,  {Fontana} A.,  {Giallongo} E.,  {Grazian} A.,    {Salimbeni} S.,
 2006, \apj, 647, 753
\bibitem[\protect\citeauthoryear{{Menci}, {Fiore}, {Puccetti} \&
 {Cavaliere}}{{Menci} et~al.}{2008}]{2008ApJ...686..219M}
{Menci} N.,  {Fiore} F.,  {Puccetti} S.,    {Cavaliere} A.,  2008, \apj, 686,
 219
 \bibitem[\protect\citeauthoryear{{Merloni} \& {Heinz}}{{Merloni} \&
 {Heinz}}{2008}]{2008MNRAS.388.1011M}
{Merloni} A.,  {Heinz} S.,  2008, \mnras, 388, 1011
\bibitem{}Mihos J.C., Hernquist L., 1994, ApJ, 425, 13
\bibitem{}Mihos J.C., Hernquist L., 1996, ApJ, 464, 641
\bibitem{}Milosavlejevic, M., Merritt D. 2001, ApJ, 563, 34
\bibitem{}Moderski R., Sikora M., 1996, MNRAS, 283, 854
\bibitem{} Monaco P., Fontanot F., 2005, MNRAS, 359, 283
\bibitem{}Mortlock D. J. et al., 2011, \nat, 474, 616
\bibitem{}Mota D.F., Van Der Bruck C., 2004, A\&A, 421, 71
\bibitem{}Mota D.F., Shaw D. J., Silk  J., 2008, ApJ, 675, 29
\bibitem{}Natarajan P., Pringle J.E., 1998, ApJ, 506, L97
\bibitem{}Netzer, H. 2009, ApJ, 695, 793 
\bibitem{}Netzer H., Trakhtenbrot B. 2007, ApJ, 654, 754 
\bibitem{}Nunes N. J., Mota D.F., 2006, MNRAS, 368, 751
\bibitem{}Omukai K., Nishi R., 1999,  ApJ, 508, 141
\bibitem{}Omukai K., 2001, ApJ, 546, 635
\bibitem{}Omukai K., Schneider R., Haiman Z., 2008, ApJ, 686, 801
\bibitem{}Pace F., Waizmann J. C., Bartelmann M., 2010, MNRAS, 406, 1865
\bibitem{}Padmanabhan T., 1993, Structure Formation in the Universe (Cambridge, UK: Cambridge University Press)
\bibitem{}Padmanabhan  J,. 2006, MNRAS, 378, 852
\bibitem{}Peacock J. A., 1999, Cosmological Physics (Cambridge: Cambridge University Press)
\bibitem{}Peebles P. J. E., 1993, Physical Cosmology (Princeton, NJ: Princeton University Press)
\bibitem{}Peebles, P. J. E., Ratra B., 1988, ApJ,  325, L17
\bibitem{}Percival W. J. et al., 2006, ApJ, 657, 645
\bibitem{}Perivolaropoulos L., 2006, Lect. Notes Phys., 720, 257
\bibitem{}Perlmutter S. et al., 1999, ApJ, 517, 565
\bibitem{}Pierre M. et al., 2011, MNRAS, 414, 1732 
\bibitem{}Pope A.C. et al., 2004 ApJ, 607, 655
\bibitem{}Predehl P. et al. 2010,
Proceedings of the conference "X-ray Astronomy 2009", Bologna, September 2009, AIPC, 1248,
arXiv:1001.2502
\bibitem{}Press W. H., Schechter P., 1974, ApJ, 187, 425
\bibitem{}Reed D., Bower R., Frenk C. S., Jenkins A., Theuns T., 2007, MNRAS, 374, 2
\bibitem{}Rees M. J., Ostriker J. P., 1977 MNRAS 179 541–59
%\bibitem{}Refregier A., Bacon D., 2003, MNRAS, 338, 48
\bibitem{}Refregier A., 2003, ARA\&A, 41, 645
%\bibitem{}Richards G. T. et al, 2006, ApJS, 166, 470
\bibitem{}Richstone D. et al., 1998, \nat, 395, A14
\bibitem{}Riess A. G. et al., 1998, AJ, 116, 1009
\bibitem{}Sahni V., Starobinsky A. A., 2000, Int. J. Mod. Phys.D, 9, 373
\bibitem{}Sahni V., Wang L.M., 2000, Phys. Rev. D, 62, 103517
\bibitem{}Sanders, D.B. et al. 1988, ApJ, 325, 74S
\bibitem{}Sethi S., Haiman Z., Pandey K., 2010, ApJ, 721, 615
\bibitem{}Shapiro S.L., 2005, ApJ, 620, 59
\bibitem{}Shankar F. et al., 2008, MNRAS, 354, 1020
\bibitem{}Scheuer P.A.G., Feiler R., 1996, MNRAS, 282, 291
\bibitem{}Shemmer O. et al. 2004, ApJ, 614, 547 
\bibitem{}Sikora M., Shwarz L., Lasota  J.-P., 2007, ApJ, 658, 815
\bibitem{}Smoot, G.F. et al. 1992, ApJ, 396, 1
\bibitem{}Soltan  A., 1982, MNRAS, 360, 565
%\bibitem{}Somerville R. S. et al., 2004b, ApJ, 600, L135
\bibitem{}Spergel D. N. et al., 2007, \apjs, 170, 377
\bibitem{}Springel V., Di Matteo T., Hernquist, L., 2005a, ApJ, 620, L79
\bibitem{}Springel V., Di Matteo T., Hernquist, L., 2005b, MNRAS, 361, 776
\bibitem{}Springel V. et al., 2005, \nat, 435, 629
\bibitem{}Stewart, K.R. 2008, ApJ, 683, 597
\bibitem{}Tegmark M. et al.,  2006, Phys. Rev. D., 74, 123507
\bibitem{}Treister E., Urry C.M., 2006, ApJ, 652, L79
\bibitem{}Treister E. et al., 2010, Science, 328, 600
\bibitem{}Trakhtenbrot, B. et al. 2011, ApJ, 730, 7 
\bibitem{}Turk M. J., Abel T., O’Shea B., 2009, Science, 325, 601
\bibitem{}Upadhye A., Ishak M., Steinhardt J.P., 2005, Phys. Rev. D, 72, 3501
\bibitem[\protect\citeauthoryear{{Volonteri}, {Haardt} \& {Madau}}{{Volonteri}
 et~al.}{2003}]{2003ApJ...582..559V}
{Volonteri} M.,  {Haardt} F.,    {Madau} P.,  2003, \apj, 582, 559
\bibitem{}Volonteri M., Rees M. J., 2005, ApJ, 633, 624
\bibitem{}Volonteri M. 2006, Invited Paper in Proceedings of "Relativistic Astrophysics and Cosmology - Einstein's Legacy" [astro-ph/0602630]
\bibitem{}Volonteri, M. 2010, Proceedings of the conference "Accretion and ejection in AGN: a global view" (Como, 22-26 June 2009)
\bibitem{}Wang L.M., Steinhardt J.P., 1998, ApJ, 508, 483
\bibitem{}Wang J.M., Ho L.C., McLure R. J., 2006, ApJ, 642, L111
\bibitem{}Weinberg S., 1989, Rev. Mod. Phys. 61, 1
\bibitem{}White S. D. M., Frenk C. S., 1991, ApJ, 379, 52
\bibitem[\protect\citeauthoryear{{Willott}, {McLure} \& {Jarvis}}{{Willott}
 et~al.}{2003}]{2003ApJ...587L..15W}
{Willott} C.~J.,  {McLure} R.~J.,    {Jarvis} M.~J.,  2003, \apjl, 587, L15
\bibitem{}Willott C. J. et al., 2010a, AJ, 139, 906
\bibitem{}Willott C. J. et al., 2010b, AJ, 140, 546
\bibitem[\protect\citeauthoryear{{Wyithe} \& {Loeb}}{{Wyithe} \&
 {Loeb}}{2003}]{2003ApJ...595..614W}
{Wyithe} J.~S.~B.,  {Loeb} A.,  2003, \apj, 595, 614
\bibitem{}Yoshida N., Omukai K., Henrquist L., Abel T., 2006, ApJ, 652, 6
\bibitem{}Yoshida N., Omukai K., Hernquist L., 2008, Science, 321, 669
\bibitem{}Yu, Q., Lu, Y. 2008, ApJ, 689, 732
\bibitem[\protect\citeauthoryear{{Zhao}, {Mo}, {Jing} \& {B{\"o}rner}}{{Zhao}
 et~al.}{2003}]{2003MNRAS.339...12Z}
{Zhao} D.~H.,  {Mo} H.~J.,  {Jing} Y.~P.,    {B{\"o}rner} G.,  2003, \mnras,
 339, 12
\bibitem{}Zlatev I., Wang L., Steinhardt P. J., 1999, Phys. Rev. Lett. 82, 896


\end{thebibliography}
%\nocite{*}

\end{document}